\documentclass[11pt,a4paper]{article}
\usepackage{graphicx}
\usepackage{amsfonts,amssymb,epsfig,amsmath,enumitem,mathtools, gensymb, upgreek}
\usepackage[utf8]{inputenc}
\usepackage{textcomp,setspace}
\usepackage{cite}
\usepackage{hyperref,caption,subcaption}
\usepackage{tabularx, booktabs}
\usepackage{xcolor,tikz}
\usetikzlibrary{shapes.geometric,positioning}

\renewcommand{\thefootnote}{\fnsymbol{footnote}}

\allowdisplaybreaks

\newcommand{\nn}[0]{\nonumber}

\begin{document}

\makeatletter \@addtoreset{equation}{section} \makeatother
\renewcommand{\theequation}{\thesection.\arabic{equation}}
\renewcommand{\thefootnote}{\fnsymbol{footnote}}

\begin{titlepage}
  \begin{center}
    \hfill {\tt KIAS-P17011}\\

    \vspace{2.5cm}

    {\LARGE\bf Little strings on $D_n$ orbifolds}

    \vspace{2cm}

    {\Large Joonho Kim and Kimyeong Lee}

    \vspace{0.5cm}

    \textit{School of Physics, Korea Institute for Advanced Study,\\
	85 Hoegiro, Seoul 02455, Republic of Korea.}

    \vspace{0.5cm}

    E-mails: {\tt joonhokim@kias.re.kr, klee@kias.re.kr}

  \end{center}

  \vspace{2.5cm}

\begin{abstract}
\begin{spacing}{1.2}
\normalsize
We explore two classes of 6d $\mathcal{N}=(1,0)$ little string theories obtained from type IIA/IIB NS5-branes probing $D_n$ singularities. 
Their tensor branches are described by effective gauge theories whose instanton solitons are macroscopic little strings. 
We specifically study two families of 2d $\mathcal{N}=(0,4)$ gauge theories which describe at low energy the worldsheet dynamics of the type IIA/IIB little strings. 
These gauge theories are useful to calculate the supersymmetric partition functions of the little string theories on $\mathbf{R}^4 \times T^2$. 
We establish the T-duality of the little string theories by utilizing their BPS spectra as a probe.
\end{spacing}

\end{abstract}
\end{titlepage}

\setcounter{tocdepth}{2}
\tableofcontents

\section{Introduction}
\label{sec:intro}

Little string theories (LSTs) are non-critical string theories defined in six dimensions \cite{Seiberg:1997zk,Berkooz:1997cq,Dijkgraaf:1996cv,Dijkgraaf:1996hk}. 
They arise from 10d superstring theory in the limit that the string coupling goes to zero, $g_s \rightarrow 0$, decoupling gravitational interactions. 
They exhibit several stringy properties including T-duality that identifies a pair of circle compactified LSTs. 
They depend on a scale parameter $m_s \sim(\alpha')^{-1/2}$ which determines the tension of little strings. 
They can be regarded as ``affine'' extensions of 6d superconformal field theories, 
which add an extra background tensor multiplet coupled to the little strings \cite{Bhardwaj:2015oru,Bhardwaj:2015xxa}.
There are a vast number of LSTs found from various combinations of branes and/or geometric singularities in the decoupling limit $g_s \rightarrow 0$ \cite{Seiberg:1997zk,Berkooz:1997cq,Dijkgraaf:1996cv,Dijkgraaf:1996hk,Blum:1997mm,Intriligator:1997dh,Brunner:1997gf,Hanany:1997gh}. 
Broader classes of LSTs can be constructed from F-theory wrapped on non-compact Calabi-Yau threefolds \cite{Bershadsky:1996nu,Aspinwall:1996vc,Aspinwall:1997ye,Bhardwaj:2015oru}.  

In this work, we study two classes of $\mathcal{N}=(1,0)$ LSTs engineered from type IIA/IIB NS5-branes probing $D_{n\geq 4}$ singularities. 
The D-type ALF spaces can be mapped via chains of string dualities to D6-O6 or NS5-ON${}^0$ 
brane systems \cite{Sen:1997kz,Sen:1998ii,Kapustin:1998fa,Hanany:1999sj} from which one can derive the effective gauge theories.
Recall that a non-gravitational $(1,0)$ theory can have tensor, vector, and hypermultiplets.
A tensor multiplet has the 2-form and scalar fields denoted by $B_i$ and $\Phi_i$. A vector multiplet has the vector field denoted by $A_i$. 
Almost all the LSTs we consider in this paper involve the same number of vector and tensor multiplets labeled by the index $i$,
where one combination of these tensor multiplets is a non-dynamical background field. The VEV of the background scalar determines the mass scale of theories, i.e.,
$\langle\Phi_{\rm b}\rangle \sim m_s^2 \sim (\alpha')^{-1}$. Particularlly in the tensor branch, where all dynamical scalars also obtain generic non-zero VEVs,
the LST allows an effective gauge theory description whose inverse gauge couplings $1/g_i^2$ are set by $\langle\Phi_i\rangle$'s. The bosonic part of the effective action for tensor and vector multiplets takes the form of 
\begin{align}
	\label{eq:6d-action}
	S_{\text{bos}} = \int \Big( \tfrac{1}{2} a_{ij} \, d\Phi_i \wedge \star d \Phi_j + \tfrac{1}{2} a_{ij} \, H_i \wedge \star H_j - a_{ij} \, \Phi_i \, \text{tr} (F_i \wedge \star F_i) + a_{ij}\,  B_i   \, \text{tr} (F_j \wedge F_j)\Big).
\end{align}
$H_i$ and $F_i$ are the 3-form and 2-form field strengths defined as 
\begin{align}
 	H_i = dB_i +  \text{tr} \left( A_i \,dA_i - \tfrac{i}{3} (A_i \, [A_i ,A_i ]) \right),\quad
 	F_i = dA_i -\tfrac{i}{2} \, \text{tr} [A_i, A_i],
\end{align}
which are invariant under the gauge transformation $\delta A_i = D \epsilon_i$ and $\delta B_i = -  \text{tr} (\epsilon_i\, dA_i)$. We regard the action as providing the field equations by varying the two-forms $B_i$, while imposing the self-duality condition $H_i = \star H_i$ on their solutions by hand. 
Note that $\mathcal{N}=(1,0)$ multiplets are all chiral, contributing to the 1-loop anomalies. For consistency at quantum level, the 1-loop gauge anomaly needs be cancelled with the  tree-level gauge anomaly that arises from the last term in the action, i.e., 
\begin{align}
	\label{eq:GS-anomaly}
	\delta S_{\text{bos}} = - a_{ij} \int   \text{tr} (\epsilon_i \, dA_i) \wedge \text{tr} (F_j \wedge F_j).
\end{align}
 This is the Green-Schwarz anomaly cancellation mechanism \cite{Green:1984sg, Sagnotti:1992qw}, which works when the 1-loop anomaly polynomial is in the factorized form such that  $I_{\text{1-loop}} = \frac{1}{2}a_{ij } \ \text{tr} (F_i \wedge F_i ) \wedge \text{tr} (F_j \wedge F_j )$.

The symmetric matrix $a_{ij}$ specifies the Dirac pairing between various 2-form charges. The common feature of LSTs is that $a_{ij}$ has precisely one null eigenvector $n_i$ \cite{Bhardwaj:2015xxa,Bhardwaj:2015oru}. The linear combinations of the tensor multiplet fields, $B_{\rm b} = \sum n_i B_i$ and $\Phi_{\rm b} = \sum n_i \Phi_i$, have vanishing kinetic terms, as being non-dynamical background fields. The VEV of the non-dynamical scalar $\Phi_{\rm b}$ defines the mass scale of the theory, 
$\langle\Phi_{\rm b}\rangle \sim m_s^2 \sim (\alpha')^{-1}$, rather than participating into the tensor branch \cite{Bhardwaj:2015oru}. And also, the non-dynamical 2-form $B_{\rm b}$ cannot participate in the Green-Schwarz mechanism, so the gauge anomaly of one gauge node must vanish by itself \cite{Bhardwaj:2015xxa,Bhardwaj:2015oru}.

Little strings are electric/magnetic sources of 2-form tensors $B_i$ with tension proportional to $\langle \Phi_i \rangle$'s. The equations of motion for
$B_i$'s are given by
\begin{align}
	d H_i = d(\star H_i) =  \text{tr} (F_i \wedge F_i).
\end{align}
The instanton solutions of the effective gauge theories supply non-zero $\text{tr} (F_i \wedge F_i)$. They are macroscopic string configurations extended over the $\mathbf{R}^{1,1} \subset \mathbf{R}^{1,5}$ directions, whose tension is set by the effective gauge coupling $1/g_i^2 \sim \langle \Phi_i \rangle$. Their instanton numbers are measured by $k_i = \frac{1}{8\pi^2} \int_{\mathbf{R}^4} \text{tr} (F_i \wedge F_i) \in \mathbf{Z}$ whose integral is taken over the transverse $\mathbf{R}^4$ directions. They satisfy $F_i = \pm \star_4 F_i $ and $H_i = \mp \star_4 d\Phi_i$ for $k_i \gtrless 0$ in which upper/lower symbols are correlated. They are little string solutions of the LSTs. We always consider the self-dual instanton solitons ($k_i >0$) from here on.

Based on the effective gauge theory description, the low energy dynamics of little strings is governed by the non-linear sigma model \cite{Manton:1981mp}. However, the sigma model description cannot be UV-complete since its target space is the instanton moduli space, involving small instanton singularities. For certain classes of 6d gauge theories, the ADHM construction provides a prescription for obtaining a UV-complete worldsheet gauge theory from the non-linear sigma model \cite{Atiyah:1978ri}. It usually agrees with the string theory realization of instanton solitons and underlying gauge theories. The brane realization can also cover the particular case that little strings are E-strings, 
which do not carry any instanton charge, providing the UV gauge theory descriptions \cite{Kim:2014dza,Kim:2015fxa} for them.
The resulting UV gauge theory is particularly useful to compute SUSY-protected observables, such as the elliptic genera of little strings.
The brane configurations associated to our LSTs will be discussed in Sections~\ref{sec:IIA}~and~\ref{sec:IIB}, from which we derive the 6d/2d gauge theories and compute the BPS partition functions.

We shall study the BPS spectrum of a circle compactified LST on $\mathbf{R}^{1,4} \times S^1$ with an Omega-deformation along the spatial $\mathbf{R}^4$ direction \cite{Nekrasov:2002qd}. Omega-deformation produces the mass gap for the $\mathbf{R}^4$ rotations, regulating the infrared divergence of LSTs.
The BPS states are the bound states between momentum and/or winding modes along the circle, which generically preserve $1/4$ SUSY generators. 
Each winding sector has a fixed winding number along the circle, which we interpret as a number of 6d little strings. 
When the circle radius is taken be large, the Hilbert space of 6d BPS states is factorized into numerous winding sectors 
being decoupled from one another at low energy regime \cite{Kim:2015gha}. Such decoupling occurs as
the ground energy gap between distinct winding sectors is proportional to the circle radius $R$, dominating the momentum excitation
proportional to the inverse of the circle radius $R^{-1}$.
The BPS spectrum of an individual winding sector is captured by the elliptic genus of worldsheet UV gauge theory,
describing the instanton strings of the 6d effective gauge theory \eqref{eq:6d-action}.
The complete 6d BPS partition function is thus constructed as the weighted sum over the 2d elliptic genera of little strings,
multiplied with an extra contribution from the pure momentum sector. This is the instanton partition function of 
the 6d effective gauge theory \eqref{eq:6d-action} on Omega-deformed $\mathbf{R}^{4} \times T^2$ \cite{Nekrasov:2002qd}.

The BPS partition function provides a powerful probe to explore T-duality of LSTs.
Since it is a protected observable under the continuous deformation of the underlying theory,
it remains to be a valid expression even beyond the large radius regime.
Recall that T-duality equates a pair of circle compactified theories, whose radii of circles are related as $R_A = m_s^{-2} / R_B$, 
with an interchange of winding and momentum states. A dual pair of LSTs are therefore expected to have the same BPS partition function. 
We shall establish the T-duality relation of the BPS spectra, by comparing the BPS partition functions of LSTs 
engineered from IIA/IIB NS5-branes on $D_n$ singularities. 
As a byproduct, our result also verifies the 5d/6d dualities involving $D_n$ singularities, discussed in \cite{DelZotto:2014hpa,Ohmori:2015pia,Hayashi:2015vhy}. Similar studies on T-duality of LSTs were already made in \cite{Kim:2015gha,Hohenegger:2015btj,Hohenegger:2016eqy,Hohenegger:2016yuv} for the $(2,0)$ LSTs and the $(1,0)$ LSTs found from NS5-branes on $A_n$ singularities.

The remaining part of this paper is organized as follows. In Section~2 and~3, we derive the 6d/2d gauge theories from the brane configurations and compute the BPS partition functions on Omega-deformed $\mathbf{R}^{4} \times T^2$. In Section~4, we study T-duality relation between the LSTs in their BPS spectra. Concluding remarks are given in Section~\ref{sec:conclusion}.

\section{IIA NS5-branes on $D_n$ orbifolds}
\label{sec:IIA}

\subsection{Effective gauge theories}
\label{subsec:IIA-gauge}

\begin{figure}[!htp]
  \begin{subfigure}{0.5\linewidth}
        \includegraphics[height=3.7cm]{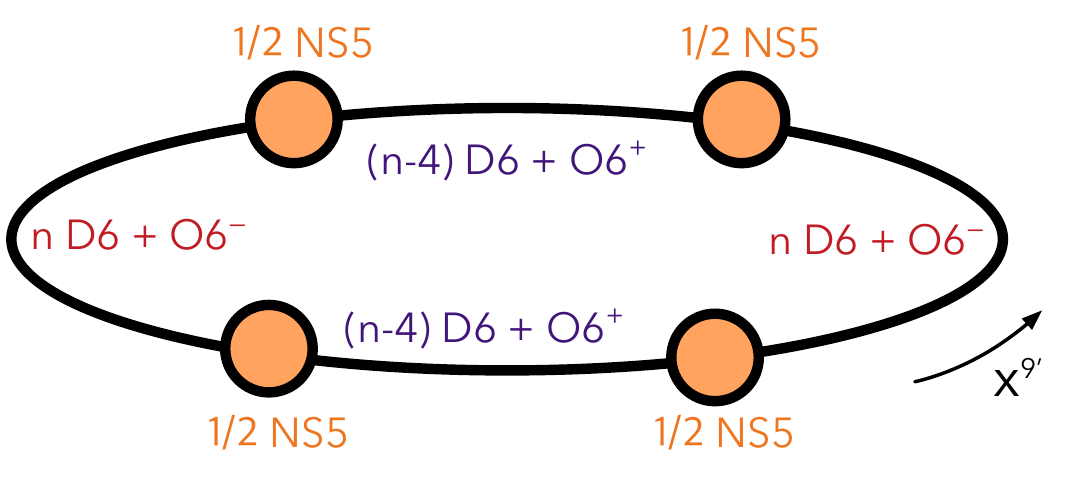}\hfill
  \end{subfigure}
  \begin{subfigure}{0.5\linewidth}
	  	\hfill\bgroup\def\arraystretch{1.6}
        \begin{tabular}{c|cccccccccc}
        	\hline\hline&0&1&2&3&4&5&6&7&8&9$'$  \\\hline
        	D6-O6 &$\bullet$&$\bullet$&$\bullet$&$\bullet$&$\bullet$&$\bullet$&&&&$\bullet$\\
        	NS5 &$\bullet$&$\bullet$&$\bullet$&$\bullet$&$\bullet$&$\bullet$&&&&\\
        	D2 &$\bullet$&$\bullet$&&&&&&&&$\bullet$\\\hline\hline
        \end{tabular}
        \egroup
  \end{subfigure}
  \caption{NS5-D6-O6 brane system at $N=2$.}
  \label{fig:IIA-NS5-brane}
\end{figure}

Recall that an S-duality transformation, also known as a ``9-11 flip'', maps type IIA string theory on the $D_n$-type ALF space to the brane construction involving $n$ D6-branes on top of an O6${}^-$ plane \cite{Sen:1997kz}. If  the ALF space fills out the $x^{6}, \cdots x^{9}$ directions, the D6-branes and O6${}^-$ plane are extended over the $x^{0}, \cdots, x^{5},x^{9'}$ directions where $x^{9'}$ denotes the M-theory circle in the original background. We now introduce $N$ NS5-branes which span the $x^0, \cdots, x^5$ directions and intersect the O6${}^-$ plane. An NS5-brane meeting an O6-plane can split into two $\frac{1}{2}$\,NS5-branes \cite{Evans:1997hk}. Moreover, a $\frac{1}{2}$\,NS5-brane provides a discrete torsion of the Kalb-Ramond field, alternating O6${}^-$ and O6${}^+$ planes. Four D6-branes are simultaneously created or annihilated due to the RR charge conservation \cite{Evans:1997hk}. The final brane configuration is illustrated in Figure~\ref{fig:IIA-NS5-brane}.

From the brane system, we derive the effective gauge theory for the LSTs engineered from IIA NS5-branes probing the $D_n$-type ALF space. The background preserves the 6d Lorentz symmetry $SO(1,5)_{012345}$ and the $SO(3)_{789}$ global symmetry, rotating the $x^7, x^8, x^9$ directions. One can decompose $SO(1,5)_{012345} \rightarrow SO(1,1)_{01} \times SU(2)_{1L} \times SU(2)_{1R}$, where $SU(2)_{1L/1R}$ generate self-dual/anti-self-dual rotations of the four-plane spanning the $x^2, \cdots, x^5$ directions. We denote the doublet indices of $SU(2)_{1L}$, $SU(2)_{1R}$, $SU(2)_{R} \cong SO(3)_{789}$ by $\alpha, \dot{\alpha}, A$ respectively. The 32 supercharges of ten-dimensional $\mathcal{N}=(1,1)$ supersymmetry can be written as $Q^{\alpha A}_{\pm \pm}$ and $Q^{\dot{\alpha}A}_{\pm \pm}$, where the first/second subscripts express eigenvalues of $\Gamma^{01}$ and $\Gamma^{9'}$ respectively. The presence of NS5-, O6-, D6-branes imposes two SUSY projectors, $\Gamma^{012345}$ and $\Gamma^{9'}$, leaving $Q^{\alpha A}_{++}$ and $Q^{\dot{\alpha}A}_{-+}$. These surviving generators satisfy the six-dimensional $\mathcal{N}=(1,0)$ SUSY algebra that contains $SU(2)_{R} \cong SO(3)_{789}$ as R-symmetry. 

The gauge symmetry comes from $2N$ stacks of D6-branes on top of O6-planes. As each stack of O6- and D6-branes is a finite segment ending on $\frac{1}{2}\,$NS5-branes, the brane configuration engineers a six-dimensional circular quiver gauge theory with $2N$ nodes. There are two types of gauge nodes. First, $n$ D6-branes plus an O6${}^-$ plane induce an $SO(2n)$ gauge symmetry. Second, $(n-4)$ D6-branes plus an O6${}^+$ plane induce an $Sp(n-4)$ gauge symmetry. The total gauge group is therefore an alternating product of $SO(2n)$ and $Sp(n-4)$ having $2N$ nodes. We label the $SO(2n)$ nodes by odd integers and the $Sp(n-4)$ nodes by even integers.

\begin{figure}[!t]
  \begin{subfigure}{0.5\linewidth}
      \centering
      \vspace{0.5cm}
        \includegraphics[height=5.5cm]{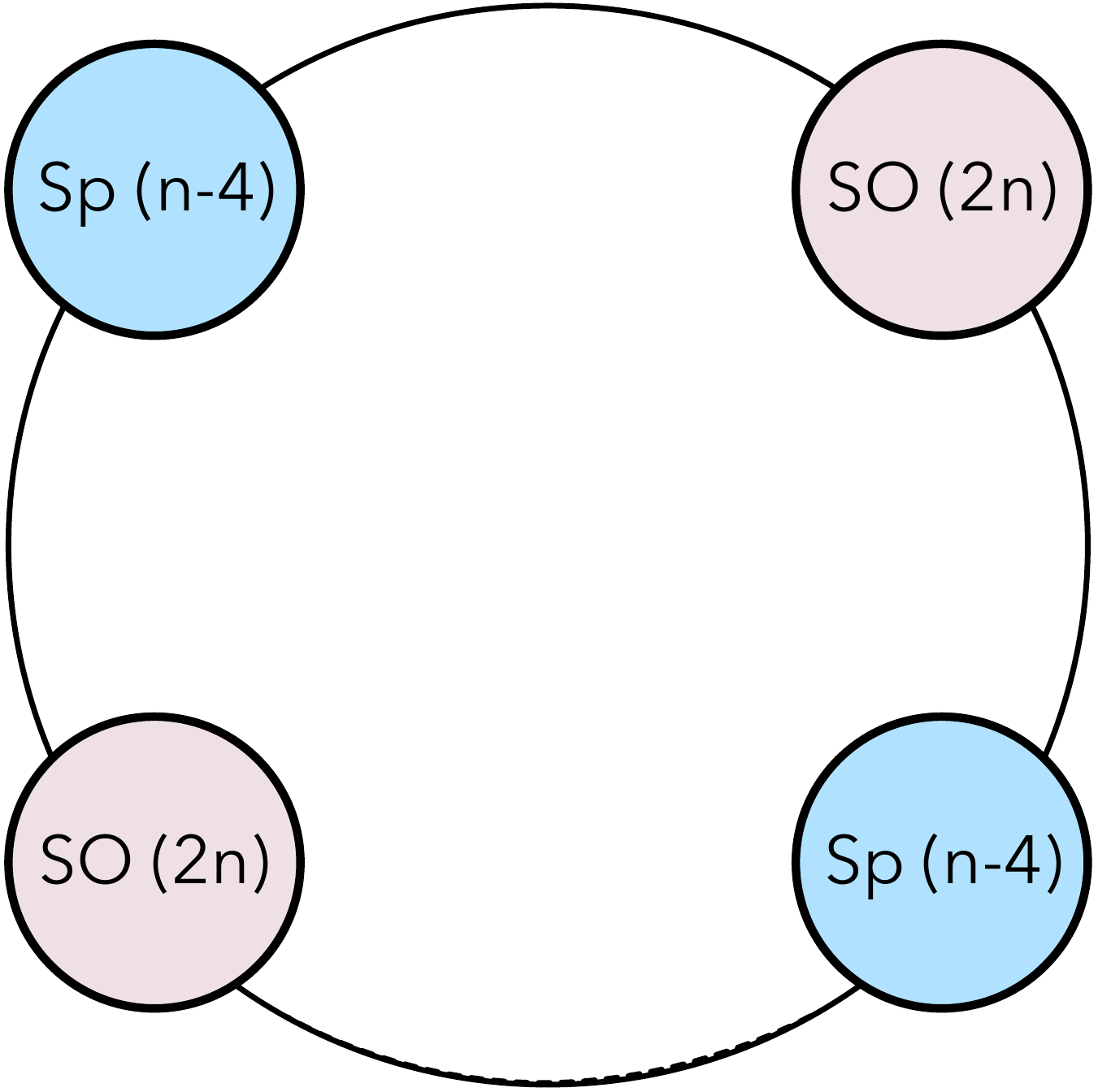}
        \vspace{0.5cm}
        \caption{6d}
        \label{fig:IIA-6d-quiver}
  \end{subfigure}
    \begin{subfigure}{0.5\linewidth}
      \centering
        \includegraphics[height=6.5cm]{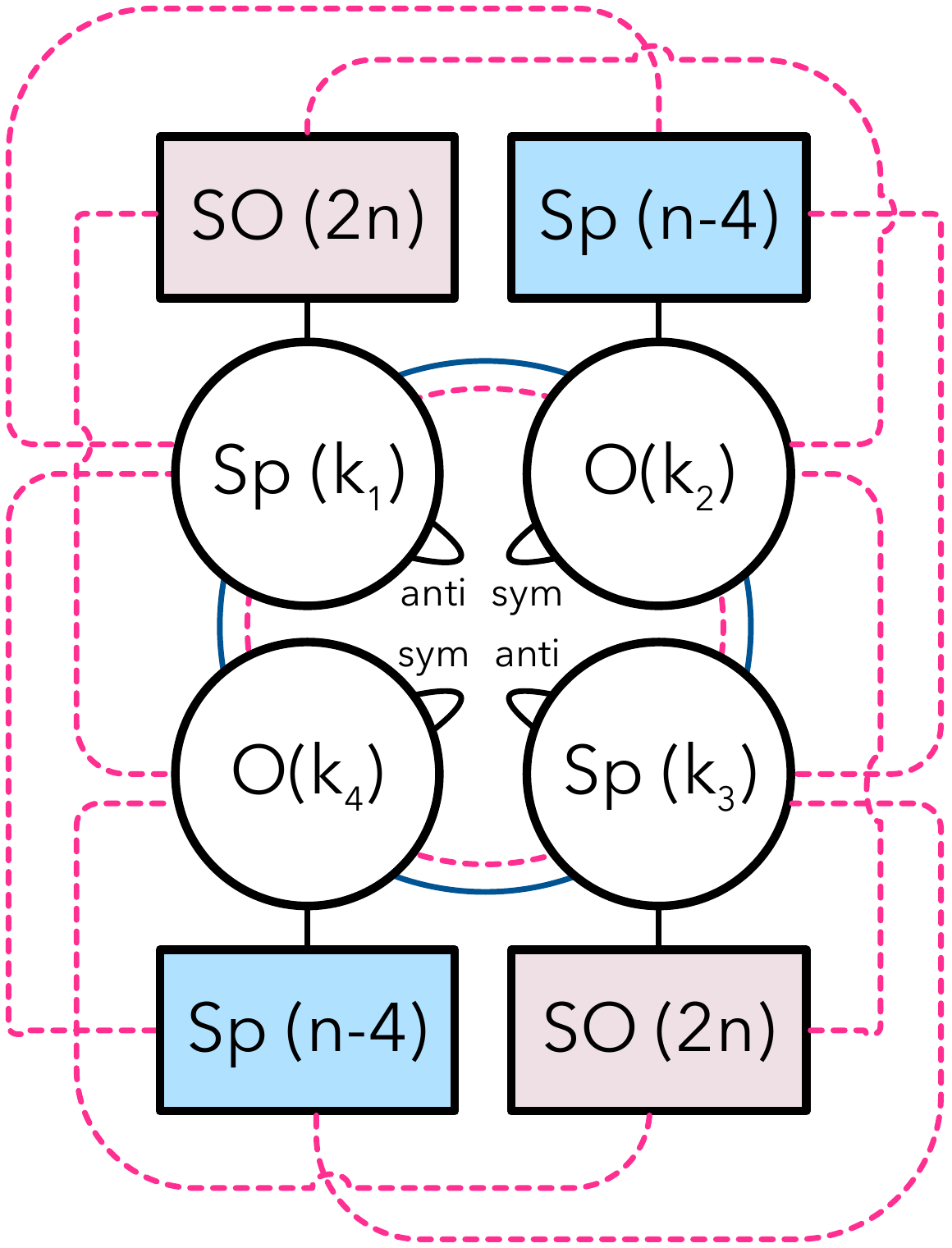}
        \caption{2d}
        \label{fig:IIA-2d-quiver}
  \end{subfigure}
  \caption{Quiver diagrams for 6d/2d gauge theories on D6/D2-branes at $N=2$.}
  \label{fig:IIA-quiver}
\end{figure}

Open strings connecting various D6-branes have massless excitation modes, corresponding to field contents in the circular quiver gauge theory. Each gauge node contains an adjoint vector multiplet. Its bosonic action, coupled to $2N$ tensor multiplets, takes the form of \eqref{eq:6d-action} with
\begin{align}
a_{ij} = \begin{dcases}
	+4 & \text{if }i=j = \text{(odd)}\\
	+1 & \text{if }i=j = \text{(even)}\\
	-1 & \text{if }i=j \pm 1\\
  0 & \text{otherwise}\\
\end{dcases}\quad\text{for }  N>1, \qquad a_{ij} = \begin{pmatrix}
	+4 & -2 \\ -2 & +1
\end{pmatrix}  \quad\text{for } N=1.
\end{align}
where $i,j \in \mathbf{Z} \text{ (mod $2N$)}$.
Each adjacent pair of gauge nodes is connected by a bifundamental half-hypermultiplet. A half-hypermultiplet is a hypermultiplet in a pseudo-real representation, whose number of fields is halved by the reality condition. Note that a half-hypermultiplet is always massless since a mass term is incompatible with the reality condition. The quiver diagram for the effective gauge theory is given in Figure~\ref{fig:IIA-quiver}\subref{fig:IIA-6d-quiver}, in which a solid line represents a half-hypermultiplet. The 6d gauge anomaly is cancelled by the Green-Schwarz mechanism \cite{Green:1984sg,Sagnotti:1992qw}.

Little string configurations are introduced in the brane system as an array of $2N$ D2-brane stacks, occupying the $x^0, x^1, x^{9'}$ directions. Each stack is a finite segment along the $x^{9'}$ direction connecting an adjacent pair of $\frac{1}{2}\,$NS5-branes. For example, the $i$-th D2-brane stack realizes instanton strings in the $i$-th gauge node, which are also fractional little strings charged under the $i$-th tensor multiplet. Every distinct configuration of little strings can be labeled by $(k_1,k_2, \cdots, k_{2N})$ where $k_i$ denotes the number of full/half D2-branes for odd/even $i$ respectively.

For the minimal case of $n=4$, no D6-branes are placed on top of the O6${}^+$ planes.
The $Sp$-type gauge symmetries as well as the bifundamental hypermultiplets become null.
However, there still exist the same $2N$ tensor multiplets and little strings which are
realized by D2-brane segments on top of O6${}^\pm$ planes. 
The brane set-up realizes a circular chain of E-string theory on O6${}^+$ and $SO(8)$ gauge theory on O6${}^-$,
where $SO(8) \times SO(8) \subset E_8$ global symmetry of E-string theory is being gauged \cite{Heckman:2013pva,DelZotto:2014hpa,Gadde2015}.

The two-dimensional gauge theory supported on the D2-branes provides the effective description for an individual winding sector in the LSTs. It inherits the $SU(2)_{1L} \times SU(2)_{1R} \times SU(2)_R$ global symmetry from the underlying 6d theory. It preserves the 4 supercharges $Q^{\dot{\alpha}A}_{-+}$, surviving after imposing an additional SUSY projector $\Gamma^{01}$ obtained from D2-branes. Note that $Q^{\dot{\alpha}A}_{-+}$ satisfies 2d $\mathcal{N}=(0,4)$ SUSY algebra which incorporates the $SO(4) \cong SU(2)_{1R} \times SU(2)_R$ as R-symmetry. It also captures the 6d gauge symmetry as flavor symmetry. 

Each stack of D2-branes supports a symplectic or an orthogonal gauge symmetry, depending on the type of orientifold plane. More precisely, $k$ full/half D2-branes on an O6${}^-$/O6${}^+$ plane support an $Sp(k)$/$O(k)$ gauge symmetry. The worldsheet gauge theory of $(k_1,k_2, \cdots, k_{2N})$ little strings is therefore an orthosymplectic circular quiver theory, whose gauge group is given by $Sp(k_1) \times O(k_2) \times \cdots \times Sp(k_{2N-1}) \times O(k_{2N})$. The field contents of the gauge theory are determined from massless modes of open strings ending on D2-branes. We summarize them as $\mathcal{N}=(0,4)$ supermultiplets in Table~\ref{tbl:IIA-matter}. The quiver diagram for the worldsheet gauge theory is also presented in Figure~\ref{fig:IIA-quiver}\subref{fig:IIA-2d-quiver}, where $\mathcal{N}=(0,4)$  hyper, twisted hyper, Fermi multiplets are denoted by black solid, blue solid, pink dashed lines. Although the theory is chiral, the field contents precisely cancel the gauge anomaly. The two-dimensional gauge anomaly is proportional to $\sum_{\psi} D_{\bf R[\psi]}$ where $\psi$ labels all chiral fermions and $\mathbf{R}[\psi]$ are their gauge representations. The index $D_{\bf R}$ is defined as $\text{Tr} (T^a_{\mathbf{R}}T^b_{\mathbf{R}}) = D_{\bf R} \delta^{ab}$.  In our cases,
\begin{align}
Sp(k_i)\text{ node}:& -4(2k_i + 2)+4(2k_i - 2)+ 4n -2(2n-8) +2(k_{i+1}+k_{i-1})- 2(k_{i+1}+k_{i-1}) = 0\\
O(k_i)\text{ node}:& -4(2k_i - 2)+4(2k_i + 2)+ 2(2n-8) - 4n +4(k_{i+1}+k_{i-1})- 4(k_{i+1}+k_{i-1}) = 0\nn
\end{align}
showing that our $\mathcal{N}=(0,4)$ gauge theory is anomaly-free. This formula also holds for $n=4$.

\begin{table}[!t]
	\textbf{\underline{Sp-type node (odd $i$):}}
	
		  	\begin{center}
		  	\bgroup\def\arraystretch{1.2}
	  		\begin{tabular}{l|l|l}
		    \hline
    Type  & Field  & Representation \\ \hline
    vector & $(A_\mu, \lambda^{\dot{\alpha}A}_-)$ & $\mathbf{adj}$ of $Sp(k_i)$               \\
    hyper & $(a_{\alpha \dot{\alpha}}, \psi^{\alpha A}_+)$ &  $\mathbf{anti}$ of $Sp(k_i)$ \\ 
    hyper & $(q_{\dot{\alpha}}, \psi_+^A)$ & $\mathbf{bif}$ of $Sp(k_i) \times SO(2n)$ \\ 
    Fermi & $(\chi_{-})_1, (\chi_-)_2$ & $\mathbf{bif}$ of $Sp(k_i) \times Sp(n-4)$  \\
    twisted hyper & $(\varphi_{\dot{\alpha}},\mu^A_+)$ &  $\mathbf{bif}$ of $Sp(k_i) \times O(k_{i+1})$  \\
    Fermi & $(\mu_{-}^\alpha)_1, (\mu_{-}^\alpha)_2$ &  $\mathbf{bif}$ of $Sp(k_i) \times O(k_{i+1})$   \\\hline
  \end{tabular}\egroup
	  	\end{center}

	  	\textbf{\underline{O-type node (even $i$):}}
	\begin{center}
		\bgroup\def\arraystretch{1.2}
		\begin{tabular}{l|l|l}
    \hline
    Type & Field & Representation \\ \hline
     vector & $(A_\mu, \lambda^{\dot{\alpha}A}_-)$ & $\mathbf{adj}$ of $O(k_i)$                \\
    hyper & $(a_{\alpha \dot{\alpha}}, \psi^{\alpha A}_+)$ &  $\mathbf{sym}$ of $O(k_i)$\\ 
     hyper & $(q_{\dot{\alpha}}, \psi_-^A)$ & $\mathbf{bif}$ of $O(k_i) \times Sp(n-4)$ \\ 
    Fermi & $(\chi_{1}), (\chi_2)$ & $\mathbf{bif}$ of $O(k_i) \times SO(2n)$ \\
     twisted hyper & $(\varphi_{\dot{\alpha}},\mu^A_+)$ &  $\mathbf{bif}$ of $O(k_i) \times Sp(k_{i+1})$ \\
   Fermi & $(\mu_{-}^\alpha)_1, (\mu_{-}^\alpha)_2$ &  $\mathbf{bif}$ of $O(k_i) \times Sp(k_{i+1})$  \\\hline
  \end{tabular}\egroup
	\end{center}
	\caption{Field contents of 2d gauge theories on D2-branes. The index $i$ is taken modulo $2N$.}
	\label{tbl:IIA-matter}
\end{table}

\subsection{BPS partition functions on $\mathbf{R}^4 \times T^2$ }
\label{subsec:IIA-index}

The 6d effective gauge theories are useful to study the BPS spectra of the LSTs on $\mathbf{R}^4 \times T^2$. For fixed $N \geq 1$ and $n\geq 4$, the partition function is defined as the following trace over the 6d Hilbert space:
\begin{align}
  \label{eq:index-iia}
  \mathcal{I}_{n,N} = \text{Tr}_{\mathcal{H}_{6d}}\,\left[ (-1)^F
q^{H_L}\bar{q}^{H_R}\, t^{J_{1R} + J_{R}} u^{J_{1L}} \prod_{i=1}^{2N}
\left(\mathfrak{n}_i^{k_i}\prod_{\ell_i=1}^{r_i}
(w_{i,\ell_i})^{F_{i,\ell_i}}\right)\right].
\end{align}
where $H_{L,R} = \frac{1}{2}(H \pm P)$ are the left/right-moving momenta along the torus $T^2$.
Using 6d $\mathcal{N}=(1,0)$ SUSY generators, $Q^{\alpha A}_{++}$ and $Q^{\dot{\alpha}A}_{-+}$,
 the right-moving Hamiltonian can be written as $H_R \sim \{Q, Q^\dagger\}$ 
where $Q \equiv Q^{\dot{1}2}_{-+}$ and $Q^\dagger \equiv -Q^{\dot{2}1}_{-+}$. 
$J_{1L}, J_{1R}, J_{R}$ are the Cartan generators for $SU(2)_{1L}$, $SU(2)_{1R}$, $SU(2)_R$ symmetries. 
$k_i$ is an instanton charge of the $i$-th gauge node, which counts the number of the $i$-th fractional little strings. 
$F_{i,\ell_i = 1,\cdots, r_i}$ are the Cartan generators of $i$-th gauge group of rank $r_i$. 
We introduce a fugacity variable for each combination of the Cartan generators that commutes with $Q$ and $Q^\dagger$. 
Besides $k_i$ and $F_{i,\ell_i= 1,\cdots, r_i}$ for $i = 1,\cdots, 2N$, there exist two more commuting combinations: $J_{1R}+J_R$, $J_{1L}$. 
The fugacity variables are also written in terms of the chemical potentials as follows.
\begin{align}
	q = e^{2\pi i \tau},\, \bar{q} = e^{2\pi i \bar{\tau}}, \, t = e^{2\pi i \epsilon_+},\, u = e^{2\pi i \epsilon_-},\, w_{i,\ell_i} = e^{2\pi i
\alpha_{i,\ell_i}}.
\end{align}
where $\tau$ is the complex parameter of the torus $T^2$. 
The background chemical potentials $\epsilon_1 = \frac{\epsilon_+ + \epsilon_-}{2}$ and $\epsilon_2 = \frac{\epsilon_+ - \epsilon_-}{2}$ 
are introduced to deform the 4-plane to the Omega-deformed $\mathbf{R}^4$. 
They are IR regulators which generate an effective mass gap for the $\mathbf{R}^4$ rotations. 
The 6d gauge holonomies $\alpha_{i,\ell_i}$ fractionalize the circle momenta. 

The 6d partition function $\mathcal{I}_{n,N}$ counts the BPS states annihilated by $Q$ and $Q^\dagger$.
These BPS states carry the left-moving momenta $H_L$ and/or the winding number $k_i$. 
When one takes the large radius limit for the spatial circle $S^1$ that little strings are wrapping on,
the Hilbert space of 6d BPS states is divided into individual winding sectors decoupled from each other at low energy.
In such limit, each sector with a fixed winding number $(k_1,\cdots,k_{2N})$ acquires the ground state energy 
 bigger than the energy scale of the momenta, so the energy gap between different winding sectors also gets bigger.
Each winding sector with a definite winding number $(k_1,\cdots,k_{2N})$ is described by the 2d SUSY gauge theory 
induced from the array of $(k_1,\cdots,k_{2N})$ D2-branes. The elliptic genus $I_{n,N}^{k_1,\cdots,k_{2N}}$ of the 2d gauge theory 
therefore captures the BPS spectrum for that particular winding sector.
The full 6d BPS partition function can be written as the sum over the 2d elliptic genera 
for individual winding sectors, weighted by the string fugacities $\mathfrak{n}_i$ conjugate to the winding numbers $k_i$,
\begin{align}
\label{eq:IIA-index-factorized}
\mathcal{I}_{n,N} = I^{0}_{n,N} \cdot \left(1 +\sum_{k_1,\cdots,k_{2N}=1}^\infty\mathfrak{n}_1^{k_1}\cdots\mathfrak{n}_{2N}^{k_{2N}}\cdot I^{k_1,\cdots,k_{2N}}_{n,N}\right),
\end{align} 
with an overall multiplication by the BPS partition function $I_{n,N}^{0}$ for the pure momentum sector.

The pure momentum sector is described by the perturbative 6d gauge theory, decoupled from non-perturbative winding modes at low energy. 
The partition function $I_{n,N}^{0}$ for the pure momentum sector collects the contribution from each $(1,0)$ supermultiplet in a multiplicative way. 
It takes the form of a plethystic exponential, 
\begin{align}
	\label{eq:pert-PE}
	\textstyle I_{n,N}^{0} = \text{PE} \left[ f_{n,N}^{0} (q,t,u,w_{i,\ell_i}) \right] \equiv \exp \left(\sum_{p=1}^\infty \frac{1}{p} \cdot f_{n,N}^{0} (q,t,u,w_{i,\ell_i})	\right),
\end{align}
where the single-particle index $f_{n,N}^{0}$ is the single letter partition function \cite{Bhattacharya:2008zy}, 
defined as a trace over the operators and their derivatives saturating the BPS condition modulo 
those operators which become zero by the equations of motion. It can also be obtained from the equivariant index theorem \cite{Nekrasov:2002qd,Shadchin:2005mx}.
The letter index 
\begin{align}
\label{eq:tr-field}
\textstyle\text{Tr}_{\rm letters}\,\left[ (-1)^F
q^{H_L}\bar{q}^{H_R}\, t^{J_{1R} + J_{R}} u^{J_{1L}} \prod_{i=1}^{2N}
\prod_{\ell_i=1}^{r_i}
(w_{i,\ell_i})^{F_{i,\ell_i}}\right],
\end{align}
is a product between the $\mathbf{R}^4$ derivative factor coming from translation modes on the $\Omega$-deformed $\mathbf{R}^4$
\begin{align}
	\label{eq:pert-r4-zero-mode}
	 \frac{t^2}{(1- tu)^2 (1-tu^{-1})^2} = \frac{1}{\sin^2 {(\pi \epsilon_1)}\cdot \sin^2 {(\pi \epsilon_2)}} ,
\end{align}
and the following factors associated to respective $\mathcal{N}=(1,0)$ multiplets:
\begin{align}
	\label{eq:pert-trace}
	\text{tensor}:&\ \ \,(\mathbf{3},\mathbf{1},\mathbf{1})_+ \oplus \ \,(\mathbf{1},\mathbf{1},\mathbf{1})_+ \oplus (\mathbf{2},\mathbf{1},\mathbf{2})_- &\hspace{-0.2cm}\rightarrow &\ \ (u+u^{-1}) ( u+u^{-1} - t-t^{-1}) \textstyle\left(\sum_{n=-\infty}^{\infty} q^n\right)^+\\
	\text{vector}:&\ \ \, (\mathbf{2},\mathbf{2},\mathbf{1})_+ \oplus \ \,(\mathbf{1},\mathbf{2},\mathbf{2})_-   &\hspace{-0.2cm}\rightarrow &\ \ (t+t^{-1}) ( u+u^{-1} - t-t^{-1})\textstyle\left(\chi_{\mathbf{R}}\,(w_{i,\ell_i}) \sum_{n=-\infty}^{\infty} q^n\right)^+ \nn \\
	\text{hyper}\,: & \ 2(\mathbf{1},\mathbf{1},\mathbf{2})_+ \oplus 2 (\mathbf{2},\mathbf{1},\mathbf{1})_- &\hspace{-0.2cm}\rightarrow &\ \ 2 ( t+t^{-1} - u-u^{-1}) \textstyle\left(\chi_{\mathbf{R}}\,(w_{i,\ell_i}) \sum_{n=-\infty}^{\infty} q^n\right)^+ \nn.
\end{align}
The triples $(\mathbf{r}_{1L},\mathbf{r}_{1R},\mathbf{r}_R)$ denote the $SU(2)_{1L}$, $SU(2)_{1R}$, $SU(2)_{R}$ representations of component fields. The $\pm$ subscript denotes if a component field is bosonic or fermionic. $\chi_{\mathbf{R}}$ is the irreducible character for a gauge representation $\mathbf{R}$ of a given supermultiplet. The $+$ superscript in the parenthesis indicates that all non-BPS states carrying non-positive momentum must be discarded. In our cases, a vector multiplet is in the adjoint representation of $SO(2n)$ or $Sp(n-4)$, for which the parenthesis becomes
\begin{align}
  SO(2n): &\quad \left[  \textstyle\sum_{\ell_i<\ell_j}^n \left(  \tfrac{w_{i,\ell_i}}{w_{i,\ell_j}}+ w_{i,\ell_i} w_{i,\ell_j}+ \tfrac{q}{w_{i,\ell_i} w_{i,\ell_j}}+ \tfrac{qw_{i,\ell_j}}{w_{i,\ell_i}} \right) + nq \right]\cdot \tfrac{1}{1-q} \\
  Sp(n-4): &\quad \left[  \textstyle\sum_{\ell_i<\ell_j}^{n-4} \left(  \tfrac{w_{i,\ell_i}}{w_{i,\ell_j}}+ w_{i,\ell_i} w_{i,\ell_j}+ \tfrac{q}{w_{i,\ell_i} w_{i,\ell_j}}+ \tfrac{qw_{i,\ell_j}}{w_{i,\ell_i}} \right) + \sum_{\ell_i=1}^{n-4} \left( w_{i,\ell_i}^2 + \tfrac{q^2}{w_{i,\ell_i}^2} \right)+ (n-4)q \right]\cdot \tfrac{1}{1-q} \nn.
\end{align}
A half-hypermultiplet is in the bifundamental representation of $SO(2n) \times Sp(n-4)$ or $Sp(n-4) \times SO(2n)$ which satisfies the pseudo-reality condition. The parenthesis becomes
\begin{align}
	\label{eq:pert-IIa-hyper}
  SO(2n) \times Sp(n-4): &\quad \textstyle \tfrac{1}{2}\left[\sum_{\ell_i=1}^{n}\left( w_{i,\ell_i} + \tfrac{q}{w_{i,\ell_i}} \right)\cdot \sum_{\ell_{i+1}=1}^{n-4}\left( w_{i+1,\ell_{i+1}} + \tfrac{q}{w_{i+1,\ell_{i+1}}} \right)\right]\cdot \tfrac{1}{1-q}\\
 Sp(n-4) \times SO(2n): &\quad \textstyle \tfrac{1}{2}\left[\sum_{\ell_{i}=1}^{n-4}\left( w_{i,\ell_{i}} + \tfrac{q}{w_{i,\ell_{i}}} \right)\cdot \sum_{\ell_{i+1}=1}^{n}\left( w_{i+1,\ell_i+1} + \tfrac{q}{w_{i+1,\ell_i+1}} \right)\right]\cdot \tfrac{1}{1-q}\nn.
\end{align} 
One obtains the final expression of $f_{n,N}^{0}$ by adding up all products of \eqref{eq:pert-r4-zero-mode} and \eqref{eq:pert-trace}. \eqref{eq:pert-PE} gives $I_{n,N}^{0}$.

We now turn to an individual sector with fixed winding numbers $(k_1,\cdots,k_{2N})$. It is described by the two-dimensional gauge theory on the array of D2-branes introduced in Section~\ref{subsec:IIA-gauge}. Thus the BPS partition function $I^{k_1,\cdots,k_{2N}}_{n,N}$ for the $(k_1,\cdots,k_{2N})$ winding sector is also given by the 2d elliptic genus of the D2-brane gauge theory.
We follow \cite{Benini:2013nda,Benini:2013xpa} for computing the elliptic genera of two-dimensional gauge theories via path integral localization. The path integral of a gauge theory can be evaluated in the weak coupling regime by performing Gaussian integrations around saddle points. The saddle points are parametrized by the gauge holonomies $A_0 + \tau A_1$ on $T^2$, classified by eigenvalues of all commuting pairs of gauge group elements. For $Sp(k)$ gauge group,
\begin{align}
  A_0 + \tau A_1 = \text{diag}\, (\pm \phi_1, \pm \phi_2, \cdots, \pm \phi_k)\quad \text{where}\quad\phi_i \in \mathbb{C}/(\mathbb{Z}+\tau\mathbb{Z}).
\end{align}
$O(k)$ group allows discrete holonomies. All disconnected holonomy sectors are classified as follows.
\begin{align}
  O(1):\ \ &\{0,\ \tfrac{1}{2}, \ \tfrac{1+\tau}{2},\ \tfrac{\tau}{2}\}\\
  O(2):\ \ &\{\text{diag}(\pm \phi_1),\ \text{diag}(0,\tfrac{\tau}{2}),\ \text{diag}(\tfrac{1}{2},\tfrac{1+\tau}{2}), \ \text{diag}(0,\tfrac{1}{2}),\ \text{diag}(\tfrac{\tau}{2},\tfrac{1+\tau}{2}), \ \text{diag}(0, \tfrac{1+\tau}{2}),  \ \text{diag}(\tfrac{1}{2}, \tfrac{\tau}{2})\}\nn\\
  O(2p+1):\ \ &\{\text{diag}(\pm \phi_1, \cdots, \pm \phi_p, 0),\ \text{diag}(\pm \phi_1, \cdots, \pm \phi_{p-1}, \tfrac{1}{2}, \tfrac{1+\tau}{2},\tfrac{\tau}{2}),\ \text{diag}(\pm \phi_1, \cdots, \pm \phi_p, \tfrac{\tau}{2}),\nn \\
           &\ \, \text{diag}(\pm \phi_1, \cdots, \pm \phi_{p-1}, \tfrac{1}{2}, \tfrac{1+\tau}{2},0), \ \text{diag}(\pm \phi_1, \cdots, \pm \phi_p, \tfrac{1}{2}),\ \text{diag}(\pm \phi_1, \cdots, \pm \phi_{p-1}, \tfrac{\tau}{2}, \tfrac{1+\tau}{2},0),\nn\\
           &\ \, \text{diag}(\pm \phi_1, \cdots, \pm \phi_p, \tfrac{1+\tau}{2}),\ \text{diag}(\pm \phi_1, \cdots, \pm \phi_{p-1},0, \tfrac{\tau}{2}, \tfrac{1}{2})\}\nn \tag*{\text{for $p\geq 1$}}\\
   O(2p):\ \ &\{\text{diag}(\pm \phi_1, \cdots, \pm \phi_p),\ \text{diag}(\pm \phi_1, \cdots, \pm \phi_{p-2},0, \tfrac{1}{2}, \tfrac{1+\tau}{2},\tfrac{\tau}{2}),\ \text{diag}(\pm \phi_1, \cdots, \pm \phi_{p-1}, 0,\tfrac{\tau}{2}),\nn \\
           &\ \, \text{diag}(\pm \phi_1, \cdots, \pm \phi_{p-1}, \tfrac{1}{2}, \tfrac{1+\tau}{2}), \ \text{diag}(\pm \phi_1, \cdots, \pm \phi_{p-1}, 0,\tfrac{1}{2}),\ \text{diag}(\pm \phi_1, \cdots, \pm \phi_{p-1}, \tfrac{\tau}{2}, \tfrac{1+\tau}{2}), \nn\\
           &\ \, \text{diag}(\pm \phi_1, \cdots, \pm \phi_{p-1}, 0,\tfrac{1+\tau}{2}),\ \text{diag}(\pm \phi_1, \cdots, \pm \phi_{p-1}, \tfrac{\tau}{2}, \tfrac{1}{2})\}\nn   \tag*{\text{for $p\geq 2$}}
\end{align}
where $\phi_i \in \mathbb{C}/(\mathbb{Z}+\tau\mathbb{Z})$. We obtain the one-loop determinant $Z_\text{1-loop}$ as the result of Gaussian integrals over massive fluctuations around a saddle point. $Z_\text{1-loop}$ is obtained as the product of the one-loop determinants over various $\mathcal{N}=(0,4)$ supermultiplets, which can be written as
\begin{align}
  \label{eq:IIA-1loop-det-vector}
  Z_{\rm vector} &= \prod_{i=1}^r 2\pi \eta\, \theta_1(2\epsilon_+) \, d\phi_i  \prod_{\rho \in \mathbf{root}}\frac{\theta_1 (\rho \cdot \phi)\,\theta_1 (\rho \cdot \phi + 2\epsilon_+)  }{\eta^2}, \\
  \label{eq:IIA-1loop-det-Fermi}
  Z_{\rm Fermi} &=  \prod_{\rho \in \mathbf{rep_g}} \prod_{\kappa \in \mathbf{rep_f}}\frac{\theta_1 (\rho \cdot \phi + \kappa\cdot z)}{\eta}\\
  \label{eq:IIA-1loop-det-hyp}
  Z_{\rm hyper} &=  \prod_{\rho \in \mathbf{rep_g}} \prod_{\kappa \in \mathbf{rep_f}}\frac{\eta}{\theta_1 (+\epsilon_+ + \rho \cdot \phi + \kappa\cdot z)}, \\
  \label{eq:IIA-1loop-det-hyp'}
  Z_\text{twisted hyper} &=  \prod_{\rho \in \mathbf{rep_g}} \prod_{\kappa \in \mathbf{rep_f}}\frac{\eta}{\theta_1 (-\epsilon_+ + \rho \cdot \phi + \kappa\cdot z)}.
\end{align}
$\rho$ is the eigenvalue for the Cartan generator of the gauge symmetry in the representation $\mathbf{rep_g}$. $\kappa$ collectively denotes the eigenvalues for the Cartan generators of $SU(2)_{1L}$ global symmetry and $\left(SO(2n) \times Sp(n-4)\right)^N$ flavor symmetry in the representation $\mathbf{rep_f}$. Their conjugate chemical potentials, i.e., $\epsilon_-$ and $\alpha_{i,\ell_i}$, are collectively denoted as $z$. Note that the 1-loop determinants of real scalars and fermions involve square roots of $\theta$'s. As they are always paired, we rearranged them as $\sqrt{\theta_1  (x+y)\theta_1(-x-y)} \sim \theta_1(x+y)$ in \eqref{eq:IIA-1loop-det-vector}-\eqref{eq:IIA-1loop-det-hyp'}. After multiplying these factors, we integrate over the zero modes which are the eigenvalues $\phi_i$ of the gauge holonomies. It is the contour integral which can be done by summing all Jeffrey-Kirwan residues, as explained in \cite{Benini:2013nda,Benini:2013xpa}. We then sum over all disconnected holonomy backgrounds, divided by the Weyl group order $\prod_{i=1}^{2N} |W_i|$.
\begin{align}
I_{n,N}^{k_1,\cdots,k_{2N}} =  \sum_\text{holonomy} \frac{1}{ (2\pi i)^{\sum_{i=1}^{2N}r_i}} \frac{1}{\prod_{i=1}^{2N} |W_i|} \oint Z_{\rm 1-loop}  
\end{align}
$|W_{i}|$ is the order of Weyl group for the $i$-th gauge node in a given holonomy background. 
\begin{align}
  &|W_{Sp(k)}| = 2^k k!, &&|W_{O(2p+1)_1}| = 2^{p+1}p!, &&|W_{O(2p+1)_3}| = 2^{p+2}(p-1)!\\
  &|W_{O(2p)_0}| = 2^p p!, &&|W_{O(2p)_2}| = 2^{p+1}(p-1)!, &&|W_{O(2p)_4}| = 2^{p+2}(p-2)!\nn
\end{align}
The subscript $\iota$ in $O(k)_\iota$ denotes the number of discrete holonomies in the background.
Finally, one can obtain the full 6d BPS partition function $\mathcal{I}_{n,N}$ from $I_{n,N}^{0}$ and $I_{n,N}^{k_1,\cdots,k_{2N}}$ using \eqref{eq:IIA-index-factorized}.

\subsubsection*{\underline{Result: 1 NS5-brane on $D_n$ singularity}}
\label{subsubsec:result}

Let us specifically consider the LSTs obtained from 1 NS5-brane probing $D_n$ singularity. There are two types of fractional little strings, realized in the brane set-up as $k_1$ full D2-branes and $k_2$ half D2-branes. We study the BPS partition functions of specific winding sectors, up to $k_1 \leq 1$ and $k_2 \leq 2$ that corresponds to a fully wound D2-brane.

\paragraph{$\mathbf{(k_1,k_2) = (0,0)}$}

The BPS spectrum of the pure momentum sector is captured by the perturbative partition function of the 6d effective gauge theory. Using \eqref{eq:pert-PE}-\eqref{eq:pert-IIa-hyper}, $I^0_{n,1}$  is written as
\begin{align}
  \text{PE}\bigg[&-\frac{(1+t^2)}{(1-tu^\pm)}\cdot 
  \bigg\{ \sum_{a<b}^n   \bigg(\frac{w_{1,a}}{w_{1,b}}+ w_{1,a} w_{1,b}+ \frac{q}{w_{1,a} w_{1,b}}+ \frac{qw_{1,b}}{w_{1,a}} \bigg) + nq \bigg\}\cdot \frac{1}{1-q} \\
  &-\frac{(1+t^2)}{(1-tu^\pm)}\cdot \bigg\{ \sum_{a<b}^{n-4} \bigg(  \frac{w_{2,a}}{w_{2,b}}+ w_{2,a} w_{2,b}+ \frac{q}{w_{2,a} w_{2,b}}+ \frac{qw_{2,b}}{w_{2,a}} \bigg) + \sum_{a=1}^{n-4} \bigg( w_{2,a}^2 + \frac{q^2}{w_{2,a}^2} \bigg)+ (n-4)q \bigg\}\cdot \frac{1}{1-q} \nn\\
  & - \frac{2\,t (u + u^{-1})}{(1-tu^\pm)}\cdot \frac{q}{1-q} + \frac{2\, t}{(1-tu^\pm)}\cdot \bigg\{ \sum_{a=1}^{n}\bigg( w_{1,a} + \frac{q}{w_{1,a}} \bigg)  \cdot  \sum_{b=1}^{n-4}\bigg(w_{2,b} + \frac{q}{w_{2,b}} \bigg) \bigg\}\cdot \frac{1}{1-q}\bigg]\nn,
\end{align}
where we used the $\pm$ notation: $(1-xy^\pm) \equiv (1-xy)(1-xy^{-1})$.

\paragraph{$\mathbf{(k_1,k_2) = (0,1) \text{ and } (0,2)}$}

These winding sectors correspond to multiple E-strings (for $n=4$) or $Sp(n-4)$ instanton strings (for $n>4$). Their elliptic genera are written in \cite{Klemm:1996hh, Haghighat:2014pva, Kim:2014dza, Kim:2015fxa,Yun:2016yzw,Hayashi:2016abm}. For $k_2 = 1$,
\begin{align}
  I_{n,1}^{0,1} = -\frac{\eta^{-6}}{2\theta_1(\epsilon_+ \pm \epsilon_-)}
                  \sum_{m=1}^4 \bigg(\frac{\prod_{a=1}^n \theta_m (\pm \alpha_{1,a})}{\prod_{b=1}^{n-4} \theta_m (\epsilon_+ \pm \alpha_{2,b})} \bigg)
\end{align}
where the abbreviated notation $\theta_m (x \pm y) \equiv \theta_m (x + y)\theta_m (x - y)$ is used. For $k_2 = 2$,
\begin{align}
  I_{n,1}^{0,2} =& +\frac{\eta^{-12}}{\theta_1(\epsilon_+\pm \epsilon_-)}\Bigg[\sum_{m=1}^4 \bigg( \frac{\prod_{a=1}^n\theta_m(\pm \alpha_{1,a} - \frac{\epsilon_++\epsilon_-}{2})^2}{4\theta_1(2\epsilon_+ + 2\epsilon_-)\theta_1(-2\epsilon_-) \prod_{b = 1}^{n-4}\theta_m (\pm \epsilon_+ \pm \alpha_{2,b} - \frac{\epsilon_+ + \epsilon_-}{2} )}+(\epsilon_- \rightarrow -\epsilon_-) \bigg) \nn\\
                &+ \sum_{c = 1}^{n-4} \bigg(\frac{\prod_{a=1}^n \theta_1 (\epsilon_+ + \alpha_{2,c} \pm \alpha_{1,a})^2}{2\theta_1 ( \epsilon_+ \pm \epsilon_- \pm 2(\alpha_{2,c} -\epsilon_+)) \theta_1(2\epsilon_+ -2\alpha_{2,c})\theta_1 (2\alpha_{2,c}) \prod_{b \neq c}^{n-4} \theta_1 (2\epsilon_+ - \alpha_{2,c} \pm \alpha_{2,b}) \theta_1 (\alpha_{2,c} \pm \alpha_{2,b})} \nn \\
                &\quad\quad\quad+ (\alpha_{2,c} \rightarrow -\alpha_{2,c}) \bigg)\ +\sum_{(m,p,r)\in \mathcal{S}}\frac{\theta_m(0)\theta_m(2\epsilon_+)\prod_{a=1}^n \theta_p( \pm \alpha_{1,a})\theta_r(\pm \alpha_{1,a})}{4\theta_1 (\epsilon_+ \pm \epsilon_-) \theta_m (\epsilon_+ \pm \epsilon_-) \prod_{b=1}^{n-4} \theta_p (\epsilon_+ \pm \alpha_{2,b}) \theta_r (\epsilon_+ \pm \alpha_{2,b})}\Bigg]\nn
\end{align}
for $\mathcal{S} = \{(2,1,2),(3,1,3),(4,1,4),(2,3,4),(3,4,2),(4,2,3)\}$.

\paragraph{$\mathbf{(k_1,k_2) = (1,0)}$}

This  sector corresponds to an $SO(2n)$ instanton string. The elliptic genus is \cite{Haghighat:2014vxa}
\begin{align}
  I_{n,1}^{1,0} = \sum_{a=1}^n\left[\frac{\eta^{12}\theta_1 (2\epsilon_+ + 2\alpha_{1,q})\theta_1(4\epsilon_+ + 2\alpha_{1,q})\prod_{b=1}^{n-4}\theta_1(\pm \alpha_{2,b} \pm (\epsilon_+ + \alpha_{1,a}))}{ 2\theta_1(\epsilon_+ \pm \epsilon_-)\prod_{c \neq a}^n \theta_1(2\epsilon_+ + \alpha_{1,a} \pm \alpha_{1,c}) \theta_1 (-\alpha_{1,a} \pm \alpha_{1,c})} + (\alpha_{1,a} \rightarrow -\alpha_{1,a})\right]
\end{align}

\paragraph{$\mathbf{(k_1,k_2) = (1,1)}$}

This winding sector is described by the two-dimensional $O(1) \times Sp(1)$
gauge theory. Its elliptic genus is given as the one-dimensional contour integral
\begin{align}
 -\oint d\phi\ \frac{\eta^{9}\,\theta_1(2\epsilon_+)\theta_1(\pm 2\phi) \theta_1 (2\epsilon_+ \pm 2\phi)\prod_{b=1}^{n-4}\theta_1(\pm \phi \pm \alpha_{2,b})}{4\theta_1(\epsilon_+ \pm \epsilon_-)^2
 \prod_{a=1}^n\theta_1 (\epsilon_+ \pm \phi \pm \alpha_{1,a})}\sum_{m=1}^4 
 \frac{\prod_{a=1}^n \theta_m(\pm \alpha_{1,a})}{\prod_{b=1}^{n-4} \theta_m (\epsilon_+ \pm \alpha_{2,b})}\frac{\theta_m (+\epsilon_- \pm \phi)^2}{\theta_m(-\epsilon_+ \pm \phi)^2} .\nn
\end{align}
The Jeffrey-Kirwan residues are obtained from the simple poles at
\begin{itemize}[noitemsep,topsep=0pt]
\item $\phi = -\epsilon_+ \pm \alpha_{1,a}$\, for $a = 1, \cdots, n$
\item $\phi = +\epsilon_+ + v_p$\ \ \ \	 for $p = 1,\cdots 4$
\end{itemize}
where $(v_1,v_2,v_3,v_4) = (0,\frac{1}{2}, \frac{1+\tau}{2}, \frac{\tau}{2})$. Collecting them all, one obtains
\begin{align}
  I_{n,1}^{1,1} &= \sum_{m=1}^4 \sum_{a=1}^n \bigg( \frac{s_m \cdot \eta^6 \ \theta_1(2 \alpha_{1,a}+2 \epsilon_+) \theta_1(2 \alpha_{1,a}+4 \epsilon_+)
  \theta_m(\epsilon_- \pm  (\alpha_{1,a} +\epsilon_+))^2}{ 4 \theta_1(\epsilon_+\pm \epsilon_-)^2 \theta_m(\alpha_{1,a}+2 \epsilon _+)^2 } \\&\times\prod_{b=1}^{n-4} \frac{  \theta_1(\pm \alpha_{2,b} \pm  (\alpha_{1,a}+\epsilon_+))}{\theta_m (\epsilon_+ \pm  \alpha_{2,b})}
      \prod_{c \neq a}^n \frac{\theta_m(\pm \alpha_{1,c}) }{\theta_1(\alpha_{1,a} \pm \alpha_{1,c})\theta_1(\alpha_{1,a} \pm \alpha_{1,c}+2 \epsilon_+)} + (\alpha_{1,a} \rightarrow -\alpha_{1,a})\bigg) \nn\\&
      -\sum_{m=1}^4\bigg(\frac{\eta ^6 \theta_1(2 \epsilon _+) \theta_1(4 \epsilon_+) \prod_{b=1}^{n-4}\theta_m(\pm \alpha_{2,b} + \epsilon _+)^2}{2 \prod_{a=1}^n\theta_m(\pm \alpha_{1,a} +2 \epsilon_+) } \bigg)\nn 
\end{align}
where the sign factor $s_m$ is defined as $s_1 = -1$, $s_{2,3,4} = 1$.

\paragraph{$\mathbf{(k_1,k_2) = (2,1)}$}
This winding sector is described by the two-dimensional $O(2) \times Sp(1) $ gauge theory which allows $7$ disconnected gauge holonomies. Six of them involve $O(2)$ discrete holonomies, contributing to the elliptic genus by the following one-dimensional contour integrals.
\begin{align}
\oint d\phi\ &\bigg(\frac{\eta^3 \theta_1(2 \epsilon _+) \theta_1(\pm 2 \phi )  \theta_1(2 \epsilon_+ \pm 2 \phi )\prod_{b=1}^{n-4}\theta_1(\pm \phi \pm \alpha_{2,b}) }{8\, \theta_1(\epsilon_+\pm \epsilon _-)^3 \prod_{a=1}^n \theta_1(\epsilon_+ \pm \phi \pm \alpha_{1,a})} \nn \\ &
  \times
  \frac{ \theta_r(0)  \theta_r(2 \epsilon _+) 
  \prod_{a=1}^n\theta_m( \pm \alpha_{1,a}) \theta_p( \pm \alpha_{1,a})
}{\theta_r(\epsilon_+ \pm \epsilon_-) \prod_{b=1}^{n-4} \theta_m(\epsilon_+ \pm \alpha_{2,b}) \theta_p(\epsilon_+\pm \alpha_{2,b})} 
\frac{\theta_m(+\epsilon_-\pm \phi )^2 \theta_p(+\epsilon_-\pm \phi )^2}{ \theta_m(-\epsilon_+ \pm \phi )^2 \theta_p(-\epsilon_+ \pm \phi)^2}   \bigg)
  \end{align}
where $(m,p,r)$ takes a value in $\{(2,1,2),(3,1,3),(4,1,4),(2,3,4),(3,4,2),(4,2,3)\}$ for each holonomy sector. The Jeffrey-Kirwan residues are obtained from the simple poles at 
\begin{itemize}[noitemsep,topsep=0pt]
\item $\phi = -\epsilon_+ \pm \alpha_{1,a}$\, for $a = 1, \cdots, n$
\item $\phi = +\epsilon_+ + v_m$
\item $\phi = +\epsilon_+ + v_p$
\end{itemize}
Summing these Jeffrey-Kirwan residues over all discrete holonomy sectors, one obtains
\begin{align}
	\label{eq:21-ell-d}
  &I_{n,1}^{2,1,d} = \sum_{(m,p,r)\in \mathcal{S}} \Bigg[\frac{\theta_r (\epsilon_+ \pm \epsilon_-)\theta_1 (2\epsilon_+) \theta_1 (4\epsilon_+)}{4\theta_1 (\epsilon_+ \pm \epsilon_-)\theta_r(0) \theta_r(2\epsilon_+)} \bigg(\prod_{a=1}^n \frac{\theta_m (\pm \alpha_{1,a})}{\theta_p (2\epsilon_+ \pm \alpha_{1,a})}\prod_{b=1}^{n-4}\frac{\theta_p(\epsilon_+ \pm \alpha_{2,b})}{\theta_m(\epsilon_+ \pm \alpha_{2,b})} + (m \leftrightarrow p) \bigg)\nn\\
                  &- \sum_{a=1}^n \bigg(\frac{s_m s_p\ \theta_r(0) \theta_r (2 \epsilon_+) \theta_1 (4\epsilon_+ - 2\alpha_{1,a}) \theta_1 (2\epsilon_+ - 2\alpha_{1,a})\theta_m (\epsilon_+ \pm \epsilon_- - \alpha_{1,a})^2\theta_p (\epsilon_+ \pm \epsilon_- - \alpha_{1,a})^2  }{8 \theta_1(\epsilon_+ \pm \epsilon_-)^3 \theta_r (\epsilon_+ \pm \epsilon_- )\theta_m (2\epsilon_+ - \alpha_{1,a})^2 \theta_p (2\epsilon_+ - \alpha_{1,a})^2 }\nn\\
                   &\times \prod_{c\neq a}^n \frac{\theta_m (\pm \alpha_{1,c})\theta_p (\pm \alpha_{1,c}) } {\theta_1(\alpha_{1,a} \pm \alpha_{1,c}) \theta_1(2\epsilon_+ - \alpha_{1,a} \pm \alpha_{1,c})} \prod_{b=1}^{n-4}\frac{ \theta_1(\epsilon_+ \pm \alpha_{2,b} - \alpha_{1,a})^2 }{ \theta_a (\epsilon_+ \pm \alpha_{2,b})  \theta_b (\epsilon_+ \pm \alpha_{2,b})} + (\alpha_{1,a} \rightarrow - \alpha_{1,a})\bigg)\Bigg]
\end{align}
The remaining sector contribute to the elliptic genus by the two-dimensional contour integral
\begin{align}
  \oint d\phi_{1} d\phi_{2} &\ \bigg(\frac{\eta^3\, \theta_1(2\epsilon_+)}{2\theta_1(\epsilon_+ \pm \epsilon_-)}
  \frac{\theta_1(+\epsilon_- \pm \phi_1 \pm \phi _2)}{\theta_1(-\epsilon_+ \pm \phi_1 \pm \phi_2)}\bigg)^2 \\ &\times
  \frac{\theta_1(\pm 2\phi_1) \theta_1(2\epsilon_+ \pm 2\phi_1) \prod_{b=1}^{n-4} \theta_1 (\pm \phi_1 \pm \alpha_{2,b})}{\prod_{a=1}^n\theta_1(\epsilon_+ \pm \alpha_{1,a} \pm \phi_1)}\frac{\prod_{a=1}^n \theta_1(\pm \alpha_{1,a} \pm \phi_2)}{\theta_1(\epsilon_+ \pm \epsilon_- \pm 2\phi_2) \prod_{b=1}^{n-4} \theta_1(\epsilon_+ \pm \alpha_{2,b} \pm \phi_2)}\nn
\end{align}
whose Jeffrey-Kirwan residues come from the following list of poles.
\begin{itemize}[noitemsep,topsep=0pt]
\item $(\phi_1, \phi_2) = (-\epsilon_+ \pm \alpha_{1,a},\ -\epsilon_+ \pm
  \alpha_{2,b})$ \ for $a = 1, \cdots, n$ and $b=1,\cdots,(n-4)$
\item $(\phi_1, \phi_2) = (2\epsilon_+ + \alpha_{2,\ell},\ -\epsilon_+ -
  \alpha_{2,\ell})$ and $(2\epsilon_+ - \alpha_{2,\ell},\ -\epsilon_+ +
  \alpha_{2,\ell})$ \ for $\ell=1,\cdots,(n-4)$
 \item $(\phi_1, \phi_2) = (-\epsilon_+ \pm \alpha_{1,i}, \ -\frac{\epsilon_+ \pm
    \epsilon_-}{2} + v_p)$ \ for $i=1,\cdots,n$ and $p=1,\cdots,4$
\item $(\phi_1, \phi_2) = (\frac{3\epsilon_+ - \epsilon_-}{2}+ v_p,\
  -\frac{\epsilon_+ - \epsilon_-}{2}-v_p)$  and $(\frac{3\epsilon_+ + \epsilon_-}{2}+ v_p,\
  -\frac{\epsilon_+ + \epsilon_-}{2}-v_p)$\ for $p=1,\cdots,4$
\end{itemize}
We add up all Jeffrey-Kirwan residues, which can be written as
\begin{align}
	\label{eq:21ell-cont}
	I_{n,1}^{2,1,c} &=+\sum_{a=1}^{n}\sum_{b=1}^{n-4}\bigg(\frac{\theta_1(\epsilon_- \pm (2\epsilon_+ - \alpha_{1,a} - \alpha_{2,b}))^2 \theta_1 (\epsilon_- \pm (\alpha_{1,a} - \alpha_{2,b}))^2 \theta_1(\epsilon_+ - \alpha_{1,a} - \alpha_{2,b})^2 }{4 \theta_1 (\epsilon_+ \pm \epsilon_-)^2 \theta_1 (-3\epsilon_+ + \alpha_{1,a}+\alpha_{2,b})^2 \theta_1 (-\epsilon_+ \pm \epsilon_- +2\alpha_{2,b}) \theta_1 (3\epsilon_+ \pm \epsilon_- - 2\alpha_{2,b})}\nn\\
  \times& \frac{\theta_1 (2\epsilon_+ - 2\alpha_{1,a}) \theta_1 (4\epsilon_+ - 2\alpha_{1,a})}{\theta_1 (2\alpha_{2,b}) \theta_1 (2\epsilon_+ - 2\alpha_{2,b})}\prod_{d\neq b}^{n-4} \frac{\theta_1(\alpha_{2,d} \pm (\epsilon_+-\alpha_{1,a}))^2 }{\theta_1(\alpha_{2,b}\pm \alpha_{2,d}) \theta_1(\pm \alpha_{2,d}-\alpha _{2,b}+2 \epsilon_+)} \nn\\
  \times& \prod _{c\neq a}^n \frac{\theta_1(\pm \alpha_{1,c}-\alpha_{2,b}+\epsilon_+)^2}{\theta_1(\alpha_{1,a} \pm \alpha_{1,c}) \theta_1(-\alpha_{1,a} \pm \alpha_{1,c}+2 \epsilon_+)}   + (\alpha_{1,a} \rightarrow -\alpha_{1,a}) + (\alpha_{2,b} \rightarrow -\alpha_{2,b}) \nn\\
  +& ( \alpha_{1,a} \rightarrow -\alpha_{1,a},\ \alpha_{2,b} \rightarrow -\alpha_{2,b}) \bigg)\nn\\
    	+&\sum_{b=1}^{n-4}\bigg(\frac{ \theta_1(\epsilon_- \pm (3\epsilon_+ + 2\alpha_{2,b})) \prod_{a=1}^n \theta_1 (\epsilon_+ + \alpha_{2,b} \pm \alpha_{1,a})^2 }{4 \eta^3 \theta_1(\epsilon_+ \pm \epsilon_-)  \theta_1 (2\epsilon_+ + 2\alpha_{2,b})^2 \theta_1 (4\epsilon_+ + 2\alpha_{2,b} ) \theta_1 (2\alpha_{2,b}) \theta_1 (\epsilon_+ \pm \epsilon_- \pm 2(\epsilon_+ + \alpha_{2,b}))  }\nn\\
  	\times&\frac{1}{\prod_{d\neq b}^{n-4} \theta_1 (2\epsilon_+ \pm \alpha_{2,d} +\alpha_{2,b})\theta_1 (\pm \alpha_{2,d} -\alpha_{2,b})} \cdot \bigg(\frac{\partial}{\partial x} \frac{\theta_1 (x-2\epsilon_+)^2 \theta_1(x-2\alpha_{2,b}-2\epsilon_+)^2 }{\theta_1 (2x - 2\epsilon_+) }\nn\\
  	\times& \frac{\theta_1(2x-2\alpha_{2,b}-5\epsilon_+ \pm \epsilon_+)  \theta_1 (2x - \epsilon_+ \pm \epsilon_-)\theta_1 (2x -3\epsilon_+ - 2\alpha_{2,b} \pm \epsilon_- )}{ \prod_{a=1}^n \theta_1 (x - 2\epsilon_+ \pm \epsilon_+ \pm \alpha_{1,a} - \alpha_{2,b})} \bigg)\bigg|_{x=0} + (\alpha_{2,b} \rightarrow -\alpha_{2,b}) \bigg)\nn\\
   - & \sum_{a=1}^n \sum_{p=1}^4\bigg( \frac{\theta_1 (4\epsilon_+ - 2\alpha_{1,a}) \theta_1 (2\epsilon_+ - 2\alpha_{1,a})}{8\theta_1(\epsilon_+ \pm \epsilon_-)^2 \theta_1 (2\epsilon_+ + 2\epsilon_-) \theta_1 (2\epsilon_-)} \nn \frac{\theta_p (\frac{\epsilon_+ +3 \epsilon_-}{2} \pm (\epsilon_+ - \alpha_{1,a}))^2}{\theta_p( \frac{3\epsilon_+ +\epsilon_-}{2} \pm (\epsilon_+ - \alpha_{1,a}))^2}\prod_{d \neq b}^{n-4}\frac{ \theta_1 (\epsilon_+ - \alpha_{1,a} \pm \alpha_{2,d})^2  }{\theta_p (\frac{\epsilon_+ + \epsilon_-}{2} \pm \epsilon_+ \pm \alpha_{2,d})}\nn\\
 \times&\frac{\theta_1 (\epsilon_+ - \alpha_{1,a} \pm \alpha_{2,b})^2\theta_p (\alpha_{1,a} \pm \frac{\epsilon_+ + \epsilon_-}{2})^2}{\theta_p (\frac{\epsilon_+ + \epsilon_-}{2} \pm \epsilon_+ \pm \alpha_{2,b})}\prod_{c \neq a}^n
   \frac{ \theta_p ( \alpha_{1,c} \pm \frac{\epsilon_+ + \epsilon_-}{2})^2}{\theta_1 (\alpha_{1,a} \pm \alpha_{1,c})\theta_1 (2\epsilon_+ - \alpha_{1,a} \pm \alpha_{1,c})}\nn\\
  +&(\alpha_{1,a} \rightarrow -\alpha_{1,a}) + (\epsilon_- \rightarrow -\epsilon_-) + (\alpha_{1,a} \rightarrow -\alpha_{1,a},\ \epsilon_- \rightarrow -\epsilon_-) \bigg)\nn\\
  +&\sum_{p=1}^4 \bigg(\frac{\theta_1 (2\epsilon_+)\prod_{a=1}^n\theta_p (\frac{\epsilon_+ + \epsilon_-}{2} \pm \alpha_{1,a})^2}{8\eta^3\, \theta_1(\epsilon_+ + \epsilon_-)^2 \theta_1 (\epsilon_+ - \epsilon_-) \theta_1(2\epsilon_-) \theta_1(3\epsilon_+ + \epsilon_-) \prod_{b=1}^{n-4} \theta_p (\frac{\epsilon_+ + \epsilon_-}{2} \pm \epsilon_+ \pm \alpha_{2,b})}\nn\\
  \times &\Big(\frac{\partial}{\partial x}\frac{\theta_1 (2x + \epsilon_+ \pm \epsilon_-) \theta_1 (2x +4\epsilon_+ \pm \epsilon_+ + \epsilon_- )\theta_1 (2x + 2\epsilon_+ +2\epsilon_-)\prod_{b=1}^{n-4}\theta_p(x + \frac{3\epsilon_+ + \epsilon_-}{2} \pm \alpha_{2,b})}{\prod_{a=1}^n\theta_p (x + \frac{3\epsilon_+ + \epsilon_-}{2} \pm \epsilon_+ \pm \alpha_{1,a})}\Big)\bigg|_{x=0}\nn\\
  + & (\epsilon_- \rightarrow -\epsilon_-) \bigg)
\end{align}
The final expression of $I_{n,1}^{2,1,c}$ can be reached by taking the derivative over $x$, then set $x$ to be zero. 
Those derivatives appear in the residues of the second and fourth poles which are double poles.
The full elliptic genus is given as the sum of \eqref{eq:21-ell-d} and \eqref{eq:21ell-cont}, i.e., $I_{n,1}^{2,1}  = I_{n,1}^{2,1,c}  + I_{n,1}^{2,1,d}$.

\section{IIB NS5-branes at $D_n$ orbifolds}
\label{sec:IIB}

\subsection{Effective gauge theories}
\label{subsec:IIB-gauge}

\begin{figure}[!htp]
  \begin{subfigure}{0.5\linewidth}
        \includegraphics[height=2.8cm]{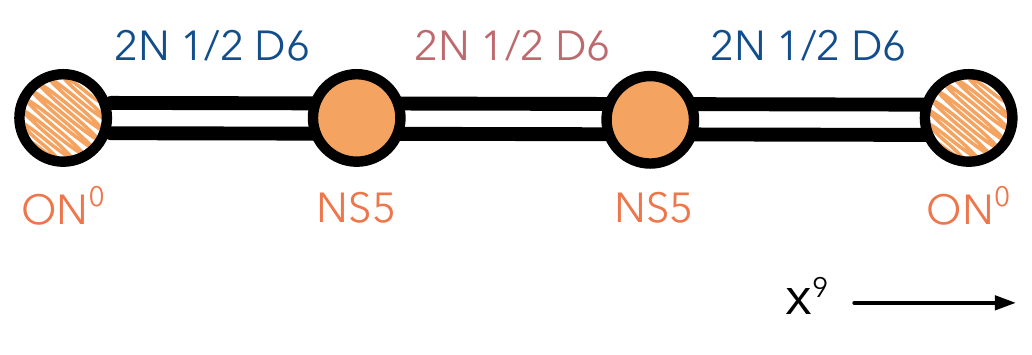}\hfill
  \end{subfigure}
  \begin{subfigure}{0.5\linewidth}
      \hfill\bgroup\def\arraystretch{1.1}
        \begin{tabular}{c|cccccccccc}
          \hline\hline&0&1&2&3&4&5&6&7&8&9  \\\hline
          D6 &$\bullet$&$\bullet$&$\bullet$&$\bullet$&$\bullet$&$\bullet$&&&&$\bullet$\\
          NS5-ON${}^0$ &$\bullet$&$\bullet$&$\bullet$&$\bullet$&$\bullet$&$\bullet$&&&&\\
          D2 &$\bullet$&$\bullet$&&&&&&&&$\bullet$\\\hline\hline
        \end{tabular}
        \egroup
  \end{subfigure}
  \caption{NS5-ON${}^0$-D6 brane system at $n=4$.}
  \label{fig:IIB-NS5-brane}
\end{figure}

We start with $N$ IIB NS5-branes which lie along the $x^0,\cdots, x^5$ directions, probing the $D_n$ ALF space in the transverse $x^6, \cdots, x^9$ directions. S-duality transformation maps the NS5-branes to D5-branes. Under T-duality transformation along the ALF circle, the D5-branes become D6-branes wrapping on the dual circle (say $x^{9''}$). The $x^{9''}$ direction is a finite segment $S^1 / \mathbb{Z}_2$ that contains an ON${}^-$ plane at each endpoint and $n$ NS5-branes in the middle \cite{Hanany:1999sj}. The fully wrapped D6-branes can be separated into various D6-brane segments ending on NS5-branes. One can obtain a weakly coupled string theory background by bringing an NS5-brane near an ON${}^-$ plane. Such chargeless combination is called an ON$^0$ plane, a perturbative string orbifold that D-branes can end on \cite{Sen:1998ii,Kapustin:1998fa,Hanany:1999sj}. We illustrate the brane configuration in Figure~\ref{fig:IIB-NS5-brane}, translating $N$ D6-branes into $2N$ half D6-branes stuck on the orbifolds.

An effective gauge theory description of the LSTs can be derived from the brane system. The brane configuration preserves the 6d Lorentz symmetry $SO(1,5)_{012345}$ and the $SO(3)_{789}$ global symmetry, rotating the $x^7, x^8, x^{9''}$ directions. We decompose $SO(1,5)_{012345} \rightarrow SO(1,1)_{01} \times SU(2)_{1L} \times SU(2)_{1R}$ and denote by $\alpha, \dot{\alpha}, A$ the doublet indices of $SU(2)_{1L}$, $SU(2)_{1R}$, $SU(2)_{R} \cong SO(3)_{789}$. The SUSY projectors imposed by NS5-branes and D6-branes are $\Gamma^{012345}$ and $\Gamma^{9''}$. As explained in Section~\ref{subsec:IIA-gauge}, $Q^{\alpha A}_{++}$ and $Q^{\dot{\alpha}A}_{-+}$ are the surviving superchanges, where the first/second subscripts denote the eigenvalues of $\Gamma^{01}$ and $\Gamma^{9''}$. They are the generators of six-dimensional $\mathcal{N}=(1,0)$ supersymmetry.  The $SU(2)_R$ global symmetry also participates into the $(1,0)$ SUSY algebra as R-symmetry.

The six-dimensional gauge symmetry comes from $(n-1)$ stacks of D6-brane segments. It is known that a stack of $2N$ half D-branes ending on an ON$^0$ plane can be described by $U(N) \times U(N)$ Yang-Mills theory without bifundamental matters \cite{Sen:1998ii,Kapustin:1998fa,Hanany:1999sj}. In our system, the leftmost and rightmost stack of D6-branes engineer four distinct $U(N)$ gauge nodes. Any other D6-brane stack engineers a $U(2N)$ gauge node. We label four $U(N)$ nodes by $i=1,\cdots,4$ and all other $U(2N)$ nodes by $i=5,\cdots, n+1$. The total gauge symmetry is therefore given by $\left(U(N)\right)^2 \times \left(U(2N)\right)^{n-3} \times \left(U(N)\right)^2$ group.

\begin{figure}[bt]
  \begin{subfigure}{0.4\linewidth}
      \centering
      \vspace{0.5cm}
        \includegraphics[height=4cm]{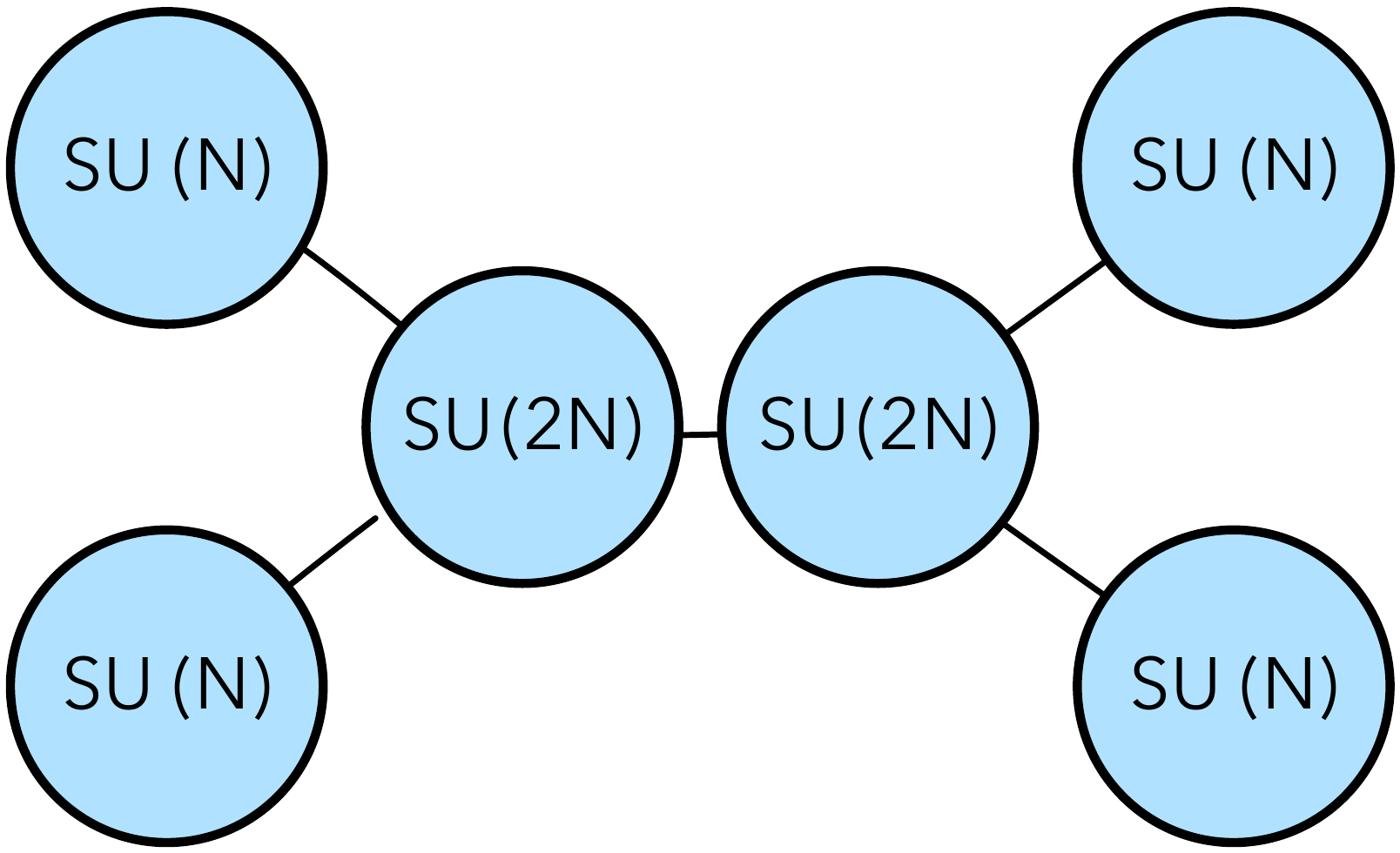}
        \vspace{0.5cm}
        \caption{6d}
        \label{fig:IIB-6d-quiver}
  \end{subfigure}
    \begin{subfigure}{0.6\linewidth}
      \hfill
        \includegraphics[height=5cm]{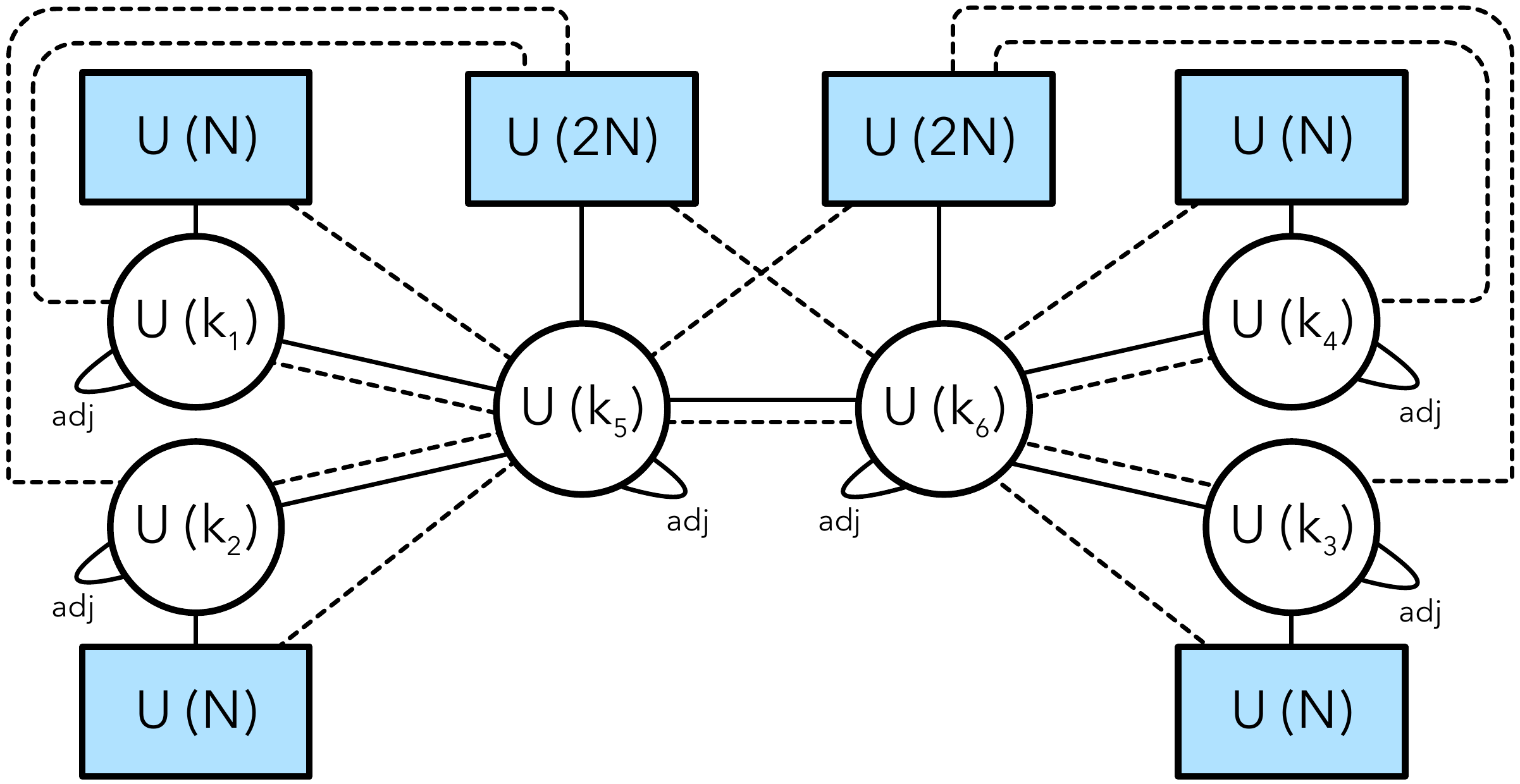}
        \caption{2d}
        \label{fig:IIB-2d-quiver}
  \end{subfigure}
	\caption{Quiver diagrams for 6d/2d gauge theories on D6/D2-branes at $n=5$.}
  \label{fig:IIB-quiver}
\end{figure}

The field contents are induced from open strings connecting various D6-branes. Each gauge node contains an adjoint vector multiplet. The bosonic Lagrangian of the vector multiplets coupled to $(n+1)$ tensor multiplets takes the form of \eqref{eq:6d-action}, whose $a_{ij}$ is given by the affine $\hat{D}_n$ Cartan matrix
\begin{align}
a_{ij} = \begin{dcases}
  +2 & \text{if }i=j\\
  -1 & \text{if }\{(i,j), (j,i)\} \cap \{(1,5),(2,5), (3,n+1),(4,n+1)\} \neq \varnothing \\
  -1 & \text{if }\{(i,j), (j,i)\} \cap \{(a,b): b = a+1  \text{ and } 5 \leq a \leq n\} \, \neq \varnothing \\
  0 &  \text{otherwise}
\end{dcases}
\end{align}
for $1\leq i,\, j \leq (n+1)$. Every pair of $i$-th and $j$-th gauge nodes, such that $a_{ij} = -1$, is connected by a massless hypermultiplet in a bifundamental representation. The quiver diagram for the effective gauge theory is therefore given by the affine $\hat{D}_n$-type Dynkin diagram, depicted in Figure~\ref{fig:IIB-quiver}\subref{fig:IIB-6d-quiver}. Our brane system realizes the Douglas-Moore construction for the $D_n$-type singularity \cite{Douglas1996}.

The Green-Schwarz mechanism cancels the non-Abelian gauge anomalies, since the 1-loop anomaly polynomial is factorized into $I_{\text{1-loop}} = \frac{1}{2}\, a_{ij} \, \text{tr} (F_i \wedge F_i) \wedge \text{tr} (F_j \wedge F_j)$ \cite{Green:1984sg, Sagnotti:1992qw}. However, the Abelian gauge symmetries also become anomalous at one-loop, i.e., 
\begin{align}
\delta S =-\epsilon_i \int (\text{tr} \, F_i) \wedge (\text{tr} \, F_i) \wedge (\text{tr}\, F_i),
\end{align}
due to those hypermultiplets charged under $(n+1)$ $U(1)$ factors. This can be cancelled through the theta term $\int \theta_i\, (\text{tr} \, F_i) \wedge (\text{tr} \, F_i) \wedge (\text{tr} \, F_i)$ where $\theta_i$ are periodic scalars induced from NS5-branes. They change under the $U(1)^{n+1}$ gauge transformation by $\delta \theta_i = \epsilon_i$. Their kinetic terms are 
written as 
\begin{align}
\int d^6x \ (\partial_\mu \theta^i - \text{tr}\, A_{\mu}^i)^2,
\end{align} 
so that the Abelian gauge fields are massive and become non-dynamical at low energy. All $U(N)$ and $U(2N)$ gauge symmetries in the effective field theory must be treated as $SU(N)$ and $SU(2N)$.

Every distinct arrangement of little strings can be labeled by $(n+1)$ integers, $(k_1, k_2, \cdots, k_{n+1})$, which correspond to the instanton charges in all gauge nodes. It is realized in the brane system as $(n+1)$ stacks of half D2-brane segments which occupy the $x^0, x^1, x^{9''}$ directions. The integer $k_i$ denotes the number of D2-branes in the $i$-th stack. From the array of D2-branes, one can derive the two-dimensional gauge theory which describes an individual winding sector of the LSTs. It inherits the $SU(2)_{1L} \times SU(2)_{1R} \times SU(2)_R$ global symmetry from the 6d theory. Introduction of the D2-branes imposes an extra SUSY projector $\Gamma^{01}$ which leaves 4 supercharges $Q^{\dot{\alpha}A}_{-+}$ unbroken. They generate 2d $\mathcal{N}=(0,4)$ supersymmetry which has $SO(4) = SU(2)_{1R} \times SU(2)_R$ as R-symmetry. 
As a stack of $k_i$ D2-branes engineers a $U(k_i)$ gauge symmetry, the total gauge group is $\prod_{i=1}^{n+1} U(k_i)$. The two-dimensional gauge theory is the affine $\hat{D}_{n}$ quiver gauge theory, which contains various field contents induced from open strings ending on D2-branes. We summarize them as $\mathcal{N}=(0,4)$ supermultiplets in Table~\ref{tbl:IIB-matter}. The Dynkin label $d_i$ is defined by $d_{1\leq i \leq 4} = 1$ and $d_{i>4} = 2$. 
We call the $i$-th and $j$-th gauge nodes are connected if $a_{ij} = a_{ji} = -1$.
See also Figure~\ref{fig:IIB-quiver}\subref{fig:IIB-2d-quiver} for the quiver diagram description. 

This theory is free of gauge anomaly, since there are the same number of left-moving and right-moving fermions in any gauge representation. The 6d gauge symmetry also appears as the flavor symmetry of the 2d gauge theory. It is the $\left(U(N)\right)^2 \times \left(U(2N)\right)^{n-3} \times \left(U(N)\right)^2$ symmetry at the classical level, but one must take into account the mixed anomalies with the gauge symmetry. Let us denote the Abelian generators of the $U(1) \subset U(k_i)$ gauge symmetry and the $U(1) \subset U(d_j N)$ flavor symmetry by $G_i$ and $F_j$. Their 2d mixed anomalies are given by
\begin{align}
\text{Tr} (\gamma_3 \,G_i F_i) = d_i N, \qquad
\text{Tr} (\gamma_3 \,G_i F_{j}) = -d_j N/2,
\end{align}
for each gauge node $i$ and its connected node $j$. Seeking for the anomaly-free $U(1)$ flavor symmetries, 
there exists the only $U(1)$ combination which has no mixed anomalies, being generated by $\sum_{i=1}^{n+1} F_i$.
However, it can be absorbed into the $U(1) \subset \prod_{i=1}^{n+1} U(k_i)$ gauge symmetry generated by $\sum_{i=1}^{n+1} G_i$.
This implies that the flavor symmetry of the 2d gauge theory must be treated as $\left(SU(N)\right)^2 \times \left(SU(2N)\right)^{n-3} \times \left(SU(N)\right)^2$ at quantum level, just as in the underlying 6d effective gauge theories.

\begin{table}[t!]
  \textbf{\underline{For each gauge node $i$:}}
          \begin{center}
        \bgroup\def\arraystretch{1.2}
\begin{tabular}{l|l|l}
    \hline
    Type  & Field  & Representation\\\hline
    vector & $(A_\mu, \lambda^{\dot{\alpha}A}_-)$ & $\mathbf{adj}$ of $U(k_i)$ \\
    hyper & $(a_{\alpha \dot{\alpha}}, \psi^{\alpha A}_+)$ &  $\mathbf{adj}$ of $U(k_i)$ \\ 
    hyper & $(q_{\dot{\alpha}}, \psi_+^A)$ & $\mathbf{bif}$ of $U(k_i) \times U(d_i N)$ \\ \hline
  \end{tabular}\egroup
      \end{center}
\vspace{0.3cm}
      \textbf{\underline{For each connected pair $(i,j)$ of gauge nodes:}}
                \begin{center}
        \bgroup\def\arraystretch{1.2}
\begin{tabular}{l|l|l}
    \hline
    Type  & Field  & Representation\\\hline
    Fermi & $(\chi_{-})_1$ & $\mathbf{bif}$ of $U(k_i) \times U(d_j  N)$\\
    Fermi & $(\chi_{-})_2$ & $\mathbf{bif}$ of $U(k_j) \times U(d_i  N)$\\
    twisted hyper & $(\varphi_{\dot{\alpha}},\mu^A_+)$ &  $\mathbf{bif}$ of $U(k_i) \times U(k_j)$\\
    Fermi & $(\mu_{-}^\alpha)_1, (\mu_{-}^\alpha)_2$ &  $\mathbf{bif}$ of $U(k_i) \times U(k_j)$ \\\hline
  \end{tabular}\egroup
      \end{center}
  \caption{Field contents of 2d gauge theories on D2-branes.}
  \label{tbl:IIB-matter}
\end{table}

\subsection{BPS partition functions on $\mathbf{R}^4 \times T^2$}
\label{subsec:IIB-index}

The BPS specta of the LSTs on Omega-deformed $\mathbf{R}^{4} \times T^2$ can be studied from the affine $\hat{D}_n$ quiver gauge theory. We define the SUSY partition function as the trace over the 6d BPS Hilbert space \cite{Nekrasov:2002qd}
\begin{align}
  \label{eq:index-iib} \mathcal{I}_{n,N} = \text{Tr}_{\mathcal{H}_{6d}}\,\left[ (-1)^F
q^{H_L}\bar{q}^{H_R}\, t^{J_{1R} + J_{R}} u^{J_{1L}} \prod_{i=1}^{n+1}
\left(\mathfrak{n}_i^{k_i}\prod_{\ell_i=1}^{k_i}
(w_{i,\ell_i})^{F_{i,\ell_i}}\right)\right].
\end{align}
Our notations for various charges and chemical potentials are already explained in Section~\ref{subsec:IIA-index}. 
Here the Cartan generators $F_{i,\ell_i}$ are those of 6d $U(N)$ and $U(2N)$ gauge symmetries which become $SU(N)$ and $SU(2N)$ at quantum level.
Their conjugate chemical potentials are therefore subject to the traceless conditions as follows.
\begin{align}
\sum_{\ell_i = 1}^{N} \alpha_{i,\ell_i} = 0 \quad\text{for $i=1,\cdots,4$} \quad\quad\text{and}\quad\quad \sum_{\ell_i = 1}^{2N} \alpha_{i,\ell_i} = 0 \quad\text{for  $i>4$}.
\end{align}
The partition function captures the 6d BPS states which carry the left-moving momenta $H_L$ along the torus $T^2$ and the winding numbers $(k_1, \cdots, k_{n+1})$.
If we take the radius $R$ of the spatial circle of the torus $T^2$ to be large, the 6d BPS Hilbert space is factorized into distinct winding sectors.
This is because the ground state energy for a winding sector is proportional to the circle radius $R$ times the winding numbers, 
dominating the energy scale $\frac{1}{R}$ of circle momenta. And also, each sector with a fixed winding number is 
described by the 2d quiver gauge theory supported on $(n+1)$ D2-brane stacks. 
We denote by $I_{n,N}^{k_1,\cdots,k_{n+1}}$ the BPS partition function of the $(k_1, \cdots, k_{n+1})$ winding sector, 
which is the elliptic genus of the 2d gauge theory on $(k_1, k_2,\cdots,k_{n+1})$ D2-brane segments.
The 6d partition function is therefore given by the weighted sum over the 2d elliptic genera, 
which captures the BPS spectrum of individual winding sectors, as follows.
\begin{align}
\label{eq:IIB-index-factorized}
  \mathcal{I}_{n,N} = I^{0}_{n,N} \cdot \left(1 + \sum_{k_1,\cdots,k_{n+1}=1}^\infty
  \mathfrak{n}_1^{k_1}\cdots\mathfrak{n}_{2N}^{k_{2N}}\cdot I^{k_1,\cdots,k_{n+1}}_{n,N}\right)
\end{align}
where the 2d elliptic genera for individual winding sectors are weighted by winding number fugacities $\mathfrak{n}_1^{k_1}\cdots\mathfrak{n}_{2N}^{k_{2N}}$. 
The overall dressing factor $I^{0}_{n,N}$ is the BPS partition function for the pure momentum sector, 
obtained from the 6d perturbative gauge theory on Omega-deformed $\mathbf{R}^4 \times T^2$.

The pure momentum sector is described by the 6d perturbative gauge theory decoupled from non-perturbative instanton modes at low energy. 
As explained in Section~\ref{subsec:IIA-index}, the partition function $I^{0}_{n,N}$ can be computed from the formula \eqref{eq:pert-PE}
where the single particle index $f^{0}_{n,N}$ is obtained from counting the letter operators \cite{Bhattacharya:2008zy}. 
Each 6d $\mathcal{N}=(1,0)$ multiplet contributes to the single particle index by the product of \eqref{eq:pert-r4-zero-mode} and \eqref{eq:pert-trace}. The parentheses in \eqref{eq:pert-trace} can be explicitly written as follows. For a vector multiplet in an adjoint representation, they are
\begin{align}
  SU(N): &\quad \left[  \textstyle\sum_{\ell_i<\ell_j}^N \left(  \tfrac{w_{i,\ell_i}}{w_{i,\ell_j}}+\tfrac{qw_{i,\ell_j}}{w_{i,\ell_i}} \right) + Nq \right] \cdot \tfrac{1}{1-q} \\
  SU(2N): &\quad \left[  \textstyle\sum_{\ell_i<\ell_j}^{2N} \left(  \tfrac{w_{i,\ell_i}}{w_{i,\ell_j}}+ \tfrac{qw_{i,\ell_j}}{w_{i,\ell_i}} \right) + 2N q \right]\cdot \tfrac{1}{1-q}\nn.
\end{align}
For a hypermultiplet in a bifundamental representation of $SU(2N) \times SU(2N)$ or $SU(N) \times SU(2N)$, 
\begin{align}
  SU(2N) \times SU(2N): &\quad \textstyle \left[\sum_{\ell_i=1}^{2N}\left( w_{i,\ell_i} + \tfrac{q}{w_{i,\ell_i}} \right)\cdot \sum_{\ell_{j}=1}^{2N}\left( w_{j,\ell_{j}} + \tfrac{q}{w_{j,\ell_{j}}} \right)\right]\cdot \tfrac{1}{1-q}\\
 SU(N) \times SU(2N): &\quad \textstyle \left[\sum_{\ell_{j}=1}^{N}\left( w_{j,\ell_{j}} + \tfrac{q}{w_{j,\ell_{j}}} \right)\cdot \sum_{\ell_j=1}^{2N}\left( w_{j,\ell_j} + \tfrac{q}{w_{j,\ell_j}} \right)\right]\cdot \tfrac{1}{1-q}.\nn
\end{align}
One can obtain the final expression of $f_{n,N}^0$ and $I_{n,N}^0$ by collecting all relevant factors and using $\eqref{eq:pert-PE}$.

An individual winding sector with fixed $(k_1, k_2,\cdots,k_{n+1})$ can be described by the two-dimensional gauge theory explained in Section~\ref{subsec:IIB-gauge}. Its BPS partition function is the elliptic genus of the gauge theory, whose computation was studied in \cite{Benini:2013nda,Benini:2013xpa} through SUSY localization. We evaluate the supersymmetric path integral in the weak coupling regime. The full path integral is reduced to Gaussian integrals around saddle points, which are parameterized by the gauge holonomy $A_0 + \tau A_1$ on $T^2$. The $U(k)$ gauge holonomy can be written as
\begin{align}
  A_0 + \tau A_1 = \text{diag}\, (\phi_1, \phi_2, \cdots, \phi_k)\quad \text{where}\quad\phi_i \in \mathbb{C}/(\mathbb{Z}+\tau\mathbb{Z}).
\end{align}
The Gaussian integration around a saddle point, labeled by eigenvalues of the $(n+1)$ gauge holonomies, results in the one-loop determinant $Z_\text{1-loop}$. It is the product of the factors \eqref{eq:IIA-1loop-det-vector}--\eqref{eq:IIA-1loop-det-hyp'} associated to each $\mathcal{N}=(0,4)$ multiplet. It can be expressed in a closed form as
 \begin{align}
   \label{eq:IIB-1-loop-det}
 	Z_\text{1-loop} &= \prod_{i=1}^{n+1}\Bigg[\left(\frac{\eta^3 \theta_1(2 \epsilon_+)}{\theta_1( \epsilon_1 )\theta_1( \epsilon_2 )} \right)^{k_i}  \prod_{p\neq q}^{k_i}\frac{\theta_1(\phi_{i,pq})\theta_1(2 \epsilon_+ + \phi_{i,pq})}{\theta_1( \epsilon_+ \pm \epsilon_- +\phi_{i,pq})} \cdot \prod_{p = 1}^{k_i} \prod_{a = 1}^{d_i N}\frac{\eta^2}{\theta_1(\epsilon_+ \pm (\phi_{i,p} - \alpha_{i,a}))}\Bigg]\\
  &\times \prod_{i \neq j}^{n+1} \left[ \prod_{p = 1}^{k_i}\prod_{a = 1}^{d_j N} \frac{\theta_1( M_{ij}\cdot (\phi_{i,p} - \alpha_{j,a}))}{\eta}  \cdot \prod_{p = 1}^{k_i} \prod_{q = 1}^{k_j} \frac{\theta_1(+\epsilon_- + M_{ij} \cdot (\phi_{i,p}-\phi_{j,q}))}{\theta_1(-\epsilon_+ +  M_{ij} \cdot (\phi_{i,p}-\phi_{j,q}))}\right] \cdot \left[\prod_{i=1}^{n+1}\prod_{p=1}^{k_i} 2\pi d \phi_{i,p}\right]\nn
 \end{align}
 where $\phi_{i,p}$ denotes the $p$-th eigenvalue of the $i$-th gauge holonomy. The following notations are also used: $M_{ij} \equiv 2\delta_{ij} - a_{ij}$ and $\phi_{i,ab} \equiv \phi_{i,a} - \phi_{i,b}$. We integrate over the eigenvalues of the gauge holonomies, $\phi_{i,p}$, parameterizing all saddle points.. It is the multi-dimensional contour integral
\begin{align}
I_{n,N}^{k_1,\cdots,k_{n+1}} = \frac{1}{ (2\pi i)^{\sum_{i=1}^{n+1}k_i}} \frac{1}{\prod_{i=1}^{n+1} k_i!} \oint Z_{\rm 1-loop}
\end{align}
which can be done by collecting the Jeffrey-Kirwan residues explained in \cite{Benini:2013nda,Benini:2013xpa}. The division factor $k_i!$ is the Weyl group order of $U(k_i)$ gauge symmetry.

The Jeffrey-Kirwan residue operation exclusively selects those poles which are classified
by colored Young diagrams. Any of selected poles, coming from the first line of
\eqref{eq:IIB-1-loop-det}, takes the following form.
\begin{align}
  \label{eq:color-young-diagram}
	\phi_{i,p} = \alpha_{i,a} -\epsilon_+ - (n_1 + n_2) \epsilon_+ - (n_1 - n_2) \epsilon_-.
\end{align}
These poles are classified by the $(d_i N)$-colored Young diagrams for all $1 \leq i\leq (n+1)$ \cite{Nekrasov:2002qd}. An $I$-colored Young diagram consists of $I$ numbers of Young tableaux. The equation \eqref{eq:color-young-diagram} indicates that the $p$-th box inside the $i$-th colored Young diagram is placed at the position $(n_1, n_2)$ of the $a$-th tableau. The rule is that a box can be piled at $(n_1, n_2)$ only if there exists a box at $(n_1 - 1, n_2)$ or $(n_1, n_2-1)$. We now claim that no additional pole can be chosen from the second line of \eqref{eq:IIB-1-loop-det}. Suppose that the values of $\phi_{i,1}, \cdots, \phi_{i,n}$ indicate $n$ boxes  inside the $i$-th colored Young diagram. If the $i$-th and $j$-th gauge nodes are interconnected, an eigenvalue $\phi_{j,1}$ of the $j$-th gauge holonomy can be determined from the poles of the following factors
\begin{align}
  \prod_{a = 1}^{d_j N} \frac{\eta}{\theta_1 (\epsilon_+ + (\phi_{j,1} - \alpha_{j,a}))}  \quad\text{ and }\quad \prod_{p=1}^n \frac{\eta}{\theta_1 (-\epsilon_+ + (\phi_{j,1} - \phi_{i,p}))}.
\end{align}
The latter one comes from the second line of \eqref{eq:IIB-1-loop-det}. Inserting back $\phi_{i,1}, \cdots, \phi_{i,n}$, this becomes
\begin{align}
  \label{eq:young-bif-bosonic}
  \prod_{a =1}^{d_i N} \prod_{(n_1,n_2) \in y_a} \frac{\eta}{\theta_1(\phi_{j,1} -\alpha_{i,a} + (n_1 +n_2)\epsilon_+  + (n_1 - n_2) \epsilon_-)}
\end{align}
where $y_a$ denotes the $a$-th tableau in the $i$-th $(d_iN)$-colored Young diagram $\mathcal{Y}_i$. Some other terms on the second line of \eqref{eq:IIB-1-loop-det} can also be written as 
\begin{align}
  \label{eq:young-bif-fermionic}
  &\prod_{a=1}^{d_i N}\frac{\theta_1 (\phi_{j,1} - \alpha_{i,a})}{\eta} \cdot \prod_{p=1}^n \frac{\theta_1 (\epsilon_- + (\phi_{i,p}-\phi_{j,1})) \theta_1 (\epsilon_- + (\phi_{j,1} - \phi_{i,p}))}{\eta^2}\\
  =& \prod_{a=1}^{d_i N}\frac{\theta_1 (\phi_{j,1} - \alpha_{i,a})}{\eta} \cdot \prod_{a =1}^{d_i N} \prod_{(n_1,n_2) \in y_a} \frac{-\theta_1 (\phi_{j,1} - \alpha_{i,a} + (n_1 + n_2 + 1) \epsilon_+ +(n_1 - n_2 \pm 1) \epsilon_- )}{\eta^2} \nn\\
  =& \prod_{a=1}^{d_i N} \Bigg[(-1)^{|y_a|} \prod_{(n_1,n_2) \in \tilde{y}_a} \frac{\theta_1 (\phi_{j,1} - \alpha_{i,a} + (n_1 + n_2 + 1) \epsilon_+ +(n_1 - n_2 \pm 1) \epsilon_- )}{\eta^2}\Bigg] \nn
\end{align}
where $\tilde{y}_a$ is the extended tableau of $y_a$ that has boxes at $(n_1,n_2)$,
$(n_1 + 1,n_2)$, $(n_1,n_2 + 1)$ for each $(n_1,n_2) \in y_a$. Since \eqref{eq:young-bif-fermionic} completely cancels out \eqref{eq:young-bif-bosonic}, the second line of  \eqref{eq:IIB-1-loop-det}  cannot develop a new pole chosen by the Jeffrey-Kirwan residue operation. It implies that every residue is associated to a distinct configuration of $(n+1)$ colored Young diagrams $\{\mathcal{Y}_1, \cdots, \mathcal{Y}_{n+1}\}$. Combining all Jeffrey-Kirwan residues, the BPS partition function $I_{n,N}^{k_1,\cdots,k_{n+1}}$ can be written as \cite{Flume:2002az,Bruzzo:2002xf} 
\begin{align}
  \label{eq:IIB-elliptic-genera}
  I_{n,N}^{k_1,\cdots,k_{n+1}}=\sum_{\{\mathcal{Y}_i\}}\prod_{i=1}^{n+1}\prod_{a=1}^{d_i N}\prod_{s \in y_a} \Bigg[&\prod_{b=1}^{d_i N}\frac{\eta^2 }{\theta_1(E_{i, ab}(s)) \theta_1 (E_{i, ab}(s)-2\epsilon_+)} \times\\ &\prod_{j \in \mathcal{N}_i}  \prod_{c=1}^{d_j N}\bigg(\frac{\theta_1 (E_{i,a}(s)-\alpha_{{j,c}})}{\eta} \prod_{s' \in y_c}\frac{\theta_1 ( E_{i,a}(s) - E_{j,c}(s') +\epsilon_-) }{\theta_1 ( E_{i,a}(s) - E_{j,c}(s') -\epsilon_+)}\bigg)\Bigg]\nn.
\end{align}
We used the following notations: $\mathcal{N}_i = \{j\ |\ a_{ij} = -1\}$. $E_{i,ab}(s)$ and $E_{i,a}(s)$ are functions of a box $s$ in the $a$-th
tableau which belongs to the $i$-th colored Young diagram, i.e.,
\begin{align}
  E_{i,ac}(s) &= \alpha_{i,a} - \alpha_{i,c} - \epsilon_+ (h_{r,a}(s) - v_{b,c} (s)-1) -
                \epsilon_- (h_{r,a} (s) + v_{b,c}(s)+1)\\
  E_{i,a}(s) &= \alpha_{i,a} - (h_{l,a}(s)+v_{t,a}(s)-1) \epsilon_+ - (h_{l,a}(s) - v_{t,a}(s)) \epsilon_-.\nn
\end{align}
$h_{l,a} (s)$ and $h_{r,a} (s)$ are the horizontal distances from the box $s$ to the leftmost and rightmost edges of $y_a$. $v_{t,a} (s)$ and $v_{t,a} (s)$ are the vertical distances from the box $s$ to the top and bottom edges of $y_a$. The 6d BPS partition function $\mathcal{I}_{n,N}$ of the LST can finally be obtained from \eqref{eq:IIB-index-factorized}.

\section{T-duality in the BPS spectra}
\label{sec:t-duality}

T-duality is a characteristic feature of LSTs that was found in many known examples. It establishes an equivalence between two circle compactified LSTs, interchanging the winding and momentum modes, when their circle radii $R$ and $\tilde{R}$ are related as $\tilde{R} = \alpha'/R$. Since IIA/IIB NS5-branes are T-dual pairs, one naturally expects that those LSTs studied in Sections~\ref{sec:IIA}~and~\ref{sec:IIB} can also be equated via T-duality.

\begin{figure}[b]
	\centering
	\includegraphics{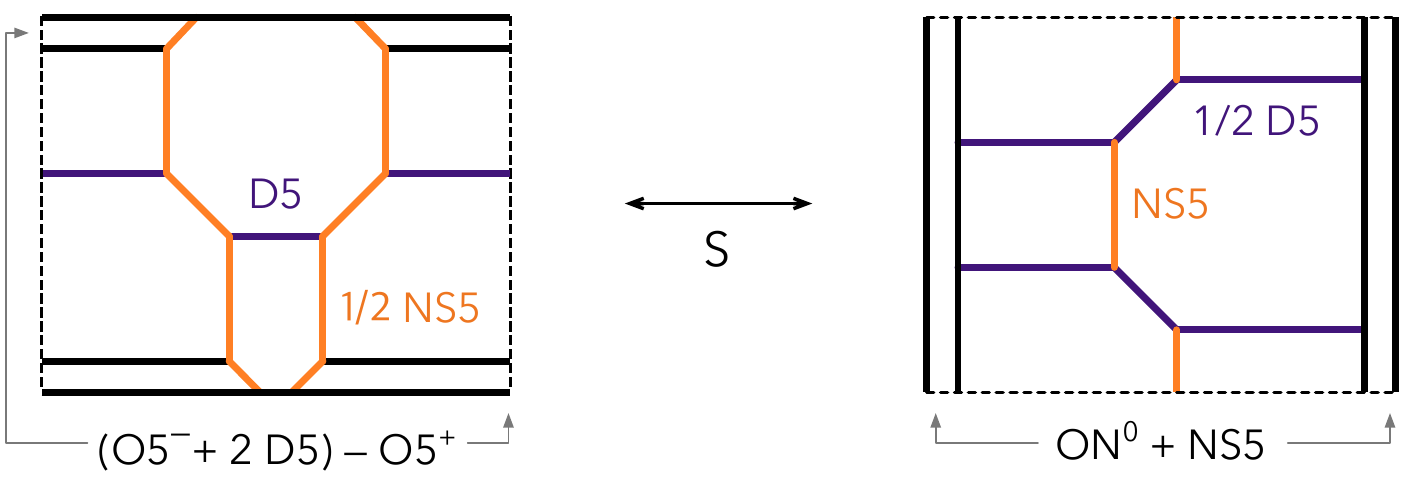}
	\caption{Duality between the generic 5-brane configurations at $N=1$ and $n=5$}
	\label{fig:brane-web-setup}
\end{figure}

A sequence of string dualities realizes T-duality of the LSTs. We start with the two brane systems that engineer the effective gauge theories studied in Sections~\ref{sec:IIA}~and~\ref{sec:IIB}. Each system contains D6-branes and NS5-branes, together with D2-brane segments corresponding to little strings. We apply T-duality transformation along the $x^1$ circle that both LSTs are wrapped on. The resulting systems are 5-brane webs. Locations of D5-branes are controlled by the chemical potentials $\alpha_{i,a}$, which were previously interpreted as the 6d gauge holonomies fractionalizing the circle momentum. Various D1- and F1-strings suspended between 5-branes are T-dual of fractional little strings and circle momenta, respectively. The equivalence of the two 5-brane web systems can be shown in two steps \cite{Hayashi:2015vhy}. First, one locates two D5-branes near the topmost O5${}^-$ planes in the D5-O5-NS5 system. This can be done by imposing the $SO(2n)$ Wilson lines,
\begin{align}
  \label{eq:wilsonline}
  2\pi A_1 = (0,0,\cdots,0,\tfrac{1}{2},\tfrac{1}{2}),
\end{align}
breaking $SO(2n) \rightarrow SO(2n-4) \times SO(4)$. Second, one takes S-duality transformation that rotates a 5-brane web diagram by 90\textdegree. It exchanges D1- and F1-strings, and D5- and NS5-branes, and the following two configurations \cite{Hayashi:2015vhy}
\begin{align}
  (\text{O5}^-  + 2\text{ D5})-\text{O5}^+ \ \longleftrightarrow \ (\text{ON}^0 + \text{NS5}).
\end{align}
More precisely, S-duality has to be performed in the singular configuration where all fractional 5-branes are unsplit \cite{Hayashi:2015vhy}. 
We move to the generic configurations afterward by separating all fractional 5-branes.
Figure~\ref{fig:brane-web-setup} depicts S-duality transformation of the two 5-brane web configurations for $N=1$ and $n=5$. 
As D1- and F1-strings correspond to winding and momentum modes of a LST, it essentially maps the winding/momentum modes of one LST to the momentum/winding modes of the other LST. This realizes the T-duality between the two LSTs, obtained from $N$ NS5-branes on the $D_n$ ALF space.

We can probe the T-duality using the BPS partition functions computed in Sections~\ref{subsec:IIA-index}~and~\ref{subsec:IIB-index}. Although the BPS partition functions are always computed in the large radius regime, they are naturally expected to be valid at any circle radius. 
For that reason, the BPS partition functions of dual LSTs must agree with each other, after suitably identifying the winding/momentum fugacities on one side with the momentum/winding fugacities on the other side. More concretely, we study if 
\begin{align}
  \mathcal{I}_{n,N}^A (q, w, \mathfrak{n}) = \mathcal{I}_{n,N}^B (q',w',\mathfrak{n}')
\end{align}
where the superscripts A/B distinguish respectively the LSTs studied in Section~\ref{sec:IIA}~and~\ref{sec:IIB}. The two sets of fugacity variables are collectively denoted as $q, w, \mathfrak{n}$ and $q',w',\mathfrak{n}'$, related by a fugacity map. The fugacity map can be best understood from the 5-brane webs diagrams as identification rules for  K\"{a}hler parameters. We derive the fugacity maps for two particular configurations: 1 NS5-brane on the $D_4$ and $D_5$ ALF spaces. From these results, one will be able to infer the maps for greater $n$-s.

\begin{figure}[b]
  \begin{subfigure}{0.5\linewidth}
        \includegraphics[height=5cm]{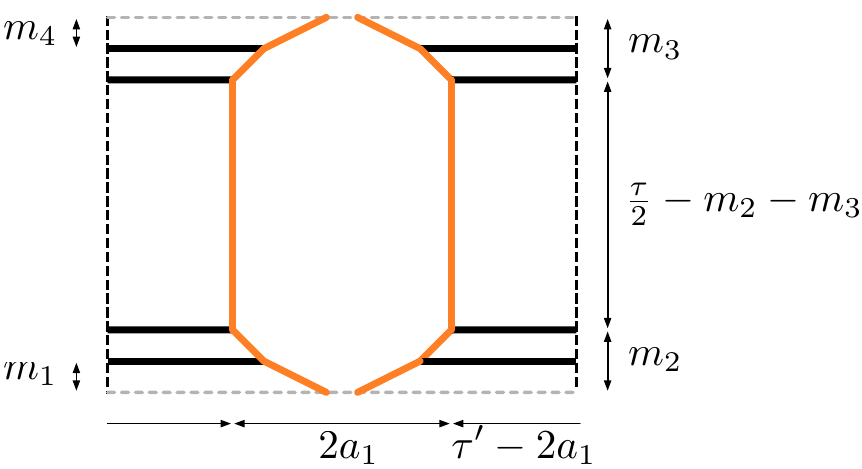}
        \caption{$N=1$ and $n=4$}
        \label{subfig:1NS5-D4}
  \end{subfigure}
  \begin{subfigure}{0.5\linewidth}
        \includegraphics[height=5cm]{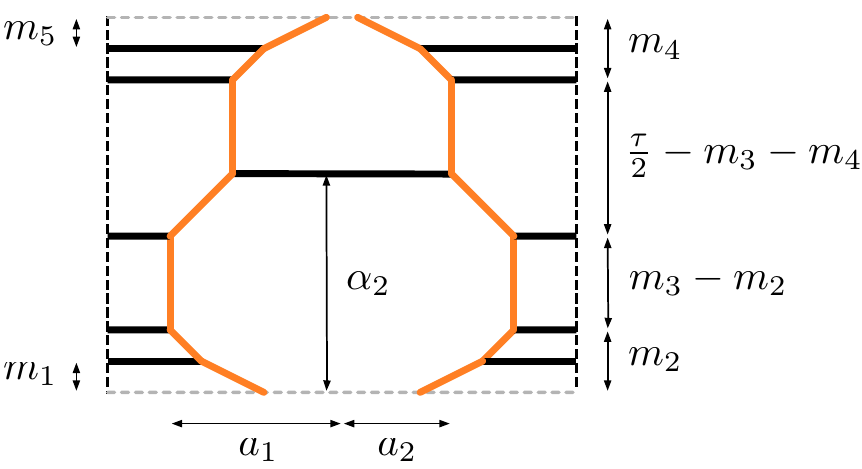}
        \caption{$N=1$ and $n=5$}
        \label{subfig:1NS5-D5}
  \end{subfigure}
  \caption{K\"{a}hler parameters in the generic 5-brane configurations at $N=1$ and $n=4,5$}
  \label{fig:1NS5-web}
\end{figure}

\paragraph{1 NS5-brane on $D_4$ singularity}
Let us identify the fugacity variables in $\mathcal{I}_{4,1}^A$ with K\"{a}hler parameters in the 5-brane web diagram of Figure~\ref{fig:1NS5-web}\subref{subfig:1NS5-D4}. There are 6 independent fugacities, $w_{1,1}, \cdots, w_{1,4}, \mathfrak{n}_1, \mathfrak{n}_2$. First, the $SO(8)$ holonomies must be replaced by
\begin{align}
w_{1,1} \rightarrow w_{1,1},\, \quad w_{1,2} \rightarrow w_{1,2},\,\quad w_{1,3} \rightarrow  q^{1/2}w_{1,3}, \,\quad  w_{1,4} \rightarrow q^{1/2}w_{1,4}
\end{align}
due to the background Wilson line \eqref{eq:wilsonline} which virtually inserts $ \exp{(\pi i\tau (F_{1,3}+F_{1,4}))}$ in the BPS partition function \eqref{eq:index-iia}. They measure the vertical locations of the D5-branes from the bottom O5-plane, so that one can identify them with K\"{a}hler parameters $m_1, \cdots, m_4, \tau$ as follows.
\begin{align}
  \label{eq:1NS-D4-Af}
  w_{1,1} = e^{2\pi i m_1} , \quad w_{1,2}= e^{2\pi i m_2}\ ,\quad q^{1/2}w_{1,3} = q^{1/2} e^{-2\pi i m_3}\ ,\quad q^{1/2}w_{1,4} = q^{1/2} e^{-2\pi i m_4}.
\end{align}
Second, the winding fugacities $\mathfrak{n}_1$ and $\mathfrak{n}_2$ measure the D1-string lengths suspended between the NS5-branes. A subtle point is that $\mathfrak{n}_2$ measures half the distance as it is conjugate to the number of half D1-strings. These fugacities are therefore identified as
\begin{align}
  \label{eq:1NS-D4-Af2}
  \mathfrak{n}_1 = e^{2\pi i (\tau' - 2a_1)}, \quad  \mathfrak{n}_2 = e^{2\pi i a_1}.
\end{align}
For example, the contribution to $\mathcal{I}_{4,1}^A$ from a fully wound D1-string is weighted with $\mathfrak{n}_1 \mathfrak{n}_2^2 = e^{2\pi i \tau'}$. 

$\mathcal{I}_{4,1}^B$ involves 6 independent variables $\mathfrak{n}_1',\, \cdots, \mathfrak{n}'_5,\, w_{5,1}'$, which we identify with the K\"{a}hler parameters in Figure~\ref{fig:1NS5-web}\subref{subfig:1NS5-D4}. First, the $SU(2)$ holonomy $w_{5,1}'$ measures the location of an NS5-brane from the middle.
\begin{align}
  \label{eq:1NS-D4-Bf}
  w_{5,1} = e^{2\pi i a_1}
\end{align}
Second, the winding fugacities $\mathfrak{n}_1',\, \cdots, \mathfrak{n}'_5$ measure the lengths of indivisible open strings, corresponding to the positive roots of affine $SO(8)$ Lie algebra. In terms of the K\"{a}hler parameters $m_1, \cdots, m_4, \tau$,
\begin{align}
  \label{eq:1NS-D4-Bf2}
  \mathfrak{n}_1'  &= e^{2\pi i (m_2 + m_1)}\ , & \mathfrak{n}_2' &= e^{2\pi i (m_2 - m_1)}\ , &&\\ \mathfrak{n}_3' &= e^{2\pi i (m_4 + m_3)}\ , & \mathfrak{n}_4' &= e^{2\pi i (m_4 - m_3)}\ , & \mathfrak{n}_5' &= e^{2\pi i (\frac{\tau}{2} - m_2 - m_3)}. \nn
\end{align}
Based on \eqref{eq:1NS-D4-Af}-\eqref{eq:1NS-D4-Bf2}, the unprimed variables are related to the primed variables as follows.
\begin{align}
  (w_{1,1},\, w_{1,2},\, w_{1,3},\, w_{1,4},\, \mathfrak{n}_1,\, \mathfrak{n}_2,\, q) = \left(\sqrt{\frac{\mathfrak{n}_1'}{\mathfrak{n}_2'}},\, \sqrt{\mathfrak{n}_1'\mathfrak{n}_2'},\, \sqrt{\frac{\mathfrak{n}_3'}{\mathfrak{n}_4'}},\, \sqrt{\mathfrak{n}_3'\mathfrak{n}_4'},\, \frac{q'}{w_{5,1}'^2},\, w_{5,1}',\, \mathfrak{n}_1'\,\mathfrak{n}_2'\,\mathfrak{n}_3'\,\mathfrak{n}_4'\,\mathfrak{n}_5'^2\right)
\end{align}
where $q = e^{2\pi i \tau}$ and $q' = e^{2\pi i \tau'}$ measure the vertical and horizontal periods in Figure~\ref{fig:1NS5-web}\subref{subfig:1NS5-D4}.

Substituting all the unprimed variables with the primed variables, we indeed found that
\begin{align}
	\mathcal{I}_{4,1}^A (q', w', \mathfrak{n}') = \mathcal{I}_{4,1}^B (q',w',\mathfrak{n}') 
\end{align}
in series expansion form, up to $q'^1 \mathfrak{n}_1^{\prime 2} \mathfrak{n}_2^{\prime 2} \mathfrak{n}_3^{\prime 2} \mathfrak{n}_4^{\prime 2} \mathfrak{n}_5^{\prime 2}$ order. The series expansion result can be written in a plethystic exponential form, i.e.,
\begin{align}
\label{eq:F41}
\mathcal{I}_{4,1}^{A/B} (q', w', \mathfrak{n}')  \equiv \text{PE} \left[\frac{t\cdot \mathcal{F}_{4,1}}{(1-tu)(1-tu^{-1})}\right],
\end{align}
where $\mathcal{F}_{4,1}$ is attached as a Mathematica file, \texttt{1NS5-D4.nb}, and also partially displayed in Appendix~\ref{sec:1ns5-D4-data}.

\paragraph{1 NS5-brane on $D_5$ singularity}

We repeat the same analysis for the $n=5$ theory, whose 5-brane web diagram is given in Figure~\ref{fig:1NS5-web}\subref{subfig:1NS5-D5}. There are 8 independent fugacity variables that appear in $\mathcal{I}_{5,1}^A$. First, the $SO(10)$ holonomies are replaced by 
\begin{align}
w_{1,1} \rightarrow w_{1,1},\, \quad w_{1,2} \rightarrow w_{1,2},\,\quad w_{1,3} \rightarrow w_{1,3},\,\quad w_{1,4} \rightarrow  q^{1/2}w_{1,4}, \,\quad  w_{1,5} \rightarrow q^{1/2}w_{1,5}
\end{align}
due to the Wilson line \eqref{eq:wilsonline} that effectively introduces $ \exp{(\pi i\tau (F_{1,4}+F_{1,5}))}$ in \eqref{eq:index-iia}. They are identified with K\"{a}hler parameters $m_1, \cdots, m_5, \tau$ as follows:
\begin{align}
  \label{eq:1NS-D5-Af}
  w_{1,1} = e^{2\pi i m_1} ,\  w_{1,2}= e^{2\pi i m_2},\ w_{1,3}= e^{2\pi i m_3},\ q^{1/2}w_{1,4} = q^{1/2} e^{-2\pi i m_4},\ q^{1/2}w_{1,5} = q^{1/2} e^{-2\pi i m_5}.
\end{align}
Second, the $Sp(1)$ fugacity $w_{2,1}$ reflects the vertical location of the middle D5-brane from the bottom O5${}^+$ plane. In terms of the K\"{a}hler parameters, it can be written as
\begin{align}
  w_{2,1} = e^{2\pi i (m_3 + a_1 - a_2)}.
\end{align}
Third, the winding fugacities $\mathfrak{n}_1$ and $\mathfrak{n}_2$ measure the lengths of D1-strings suspended between the NS5-branes. Being different from the $n=4$ case, however, two options seem available: One depends on $a_1$ while the other depends on $a_2$. We temporarily take the limit $\tau \rightarrow \infty$ to clarify this. It was observed in \cite{Bao:2013pwa,Hayashi:2013qwa} that the length of D1-strings corresponds to the average of  horizontal distances between upper/lower pairs of asymptotic NS5-branes. In our system, it implies that
\begin{align}
  \mathfrak{n}_1 = e^{2\pi i (\tau' - 2a_2)}, \quad  \mathfrak{n}_2 = e^{2\pi i a_2}
\end{align}
which must be true also for finite $\tau$ due to the robustness of the BPS spectrum.

We turn to $\mathcal{I}_{5,1}^B$ involving 8 fugacities $\mathfrak{n}_1',\, \cdots, \mathfrak{n}'_6,\, w_{5,1}',\, w_{6,1}'$. First, the $SU(2)$ holonomies $w_{5,1}'$ and $w_{6,1}'$ measure the horizontal locations of the lower and upper NS5-branes from the middle, respectively.
\begin{align}
  w_{5,1}' = e^{2\pi i a_1}\,,\quad w_{6,1}' = e^{2\pi i a_2}
\end{align}
Second, the winding fugacities $\mathfrak{n}_1',\, \cdots, \mathfrak{n}'_6$ are the lengths of indivisible open strings which correspond to the positive roots of affine $SO(10)$ Lie algebra.
\begin{align}
  \label{eq:1NS-D5-Bf2}
  \mathfrak{n}_1'  &= e^{2\pi i (m_2 + m_1)}\ , & \mathfrak{n}_2' &= e^{2\pi i (m_2 - m_1)}\ , & \mathfrak{n}_3' &= e^{2\pi i (m_5 + m_4)}\ ,\\ \mathfrak{n}_4' &= e^{2\pi i (m_5 - m_4)}\ , & \mathfrak{n}_5' &= e^{2\pi i (m_3 - m_2)}\ , & \mathfrak{n}_6' &= e^{2\pi i (\frac{\tau}{2} - m_3 - m_4)}. \nn
\end{align}
Based on \eqref{eq:1NS-D5-Af}-\eqref{eq:1NS-D5-Bf2}, the unprimed variables are related to the primed variables as follows.
\begin{align}
  (w_{1,1},\, w_{1,2},\, w_{1,3},\, w_{1,4},\, w_{1,5}) &= \left(\sqrt{\frac{\mathfrak{n}_1'}{\mathfrak{n}_2'}},\, \sqrt{\mathfrak{n}_1'\mathfrak{n}_2'},\, \sqrt{\mathfrak{n}_1'\mathfrak{n}_2'} \mathfrak{n}_5', \,\sqrt{\frac{\mathfrak{n}_3'}{\mathfrak{n}_4'}},\, \sqrt{\mathfrak{n}_3'\mathfrak{n}_4'}\right) \nn \\
  (w_{2,1}, \, \mathfrak{n}_1,\, \mathfrak{n}_2,\, q) &=  \left( \frac{\sqrt{\mathfrak{n}_1'\mathfrak{n}_2'} \mathfrak{n}_5' \cdot w_{5,1}'}{w_{6,1}'}, \frac{q'}{w_{6,1}^{\prime 2}},\, w_{6,1}',\, \mathfrak{n}_1' \mathfrak{n}_2'\mathfrak{n}_3'\mathfrak{n}_4'\mathfrak{n}_5^{\prime 2}\mathfrak{n}_6^{\prime 2}\right)
\end{align}
where $q = e^{2\pi i \tau}$ and $q' = e^{2\pi i \tau'}$ correspond to the vertical and horizontal periods in Figure~\ref{fig:1NS5-web}\subref{subfig:1NS5-D5}.

Replacing all the unprimed variables with the primed variables, we again found that
\begin{align}
  \mathcal{I}_{5,1}^A (q', w', \mathfrak{n}') = \mathcal{I}_{5,1}^B (q',w',\mathfrak{n}') 
\end{align}
in series expansion form, up to $q'^1 \mathfrak{n}_1^{\prime 1} \mathfrak{n}_2^{\prime 1} \mathfrak{n}_3^{\prime 1} \mathfrak{n}_4^{\prime 1} \mathfrak{n}_5^{\prime 2}\mathfrak{n}_6^{\prime 2}$ order. The series expansion result can be written in a plethystic exponential form, i.e.,
\begin{align}
\mathcal{I}_{5,1}^{A/B} (q', w', \mathfrak{n}')  \equiv \text{PE} \left[\frac{t\cdot \mathcal{F}_{5,1}}{(1-tu)(1-tu^{-1})}\right],
\end{align}
where the polynomial $\mathcal{F}_{5,1}$ is attached as a separate Mathematica notebook file, \texttt{1NS5-D5.nb}.

\paragraph{Generalizations} 
We state the fugacity map for general $n>5$ extending the above results. The two sets of $(2n-2)$ fugacity variables, appearing in $\mathcal{I}_{n,1}^A$ and $\mathcal{I}_{n,1}^B$, are respectively denoted as
\begin{align}
\mathcal{I}_{n,1}^A&: \mathfrak{n}_1, \, \mathfrak{n}_2, \, w_{1,1}, \, \cdots, w_{1,n}, \, w_{2,1}, \, \cdots, w_{2,n-4}\\
\mathcal{I}_{n,1}^B&: \mathfrak{n}_1',\, \cdots, \mathfrak{n}'_{n+1},\, w_{5,1}',\, \cdots, w_{n+1,1}'.
\end{align}
They can be identified with each other through K\"{a}hler parameters in the corresponding 5-brane web. 
The $SO(2n)$ holonomies $w_{1,1}, \, \cdots, w_{1,n}$ must be replaced by
\begin{align}
w_{1,n-1} \rightarrow  q^{1/2}w_{1,n-1}, \,\quad  w_{1,n} \rightarrow q^{1/2}w_{1,n}, \, \quad w_{1,a} \rightarrow w_{1,a} \quad \text{for all $a \leq (n-2)$}.
\end{align}
due to the Wilson line \eqref{eq:wilsonline} which induces $ \exp{(\pi i\tau (F_{1,n-1}+F_{1,n}))}$ in \eqref{eq:index-iia}. Repeating the same analysis, the unprimed variables can be related to the primed variables as follows.
\begin{align}
  (w_{1,1},\, w_{1,2},\,  w_{1,n-1},\, w_{1,n}) &= \left(\sqrt{\frac{\mathfrak{n}_1'}{\mathfrak{n}_2'}},\, \sqrt{\mathfrak{n}_1'\mathfrak{n}_2'},\,\sqrt{\frac{\mathfrak{n}_3'}{\mathfrak{n}_4'}},\, \sqrt{\mathfrak{n}_3'\mathfrak{n}_4'}\right) \nn \\
  (w_{1,3},\, \cdots ,\, w_{1,n-2}) &= \left(\sqrt{\mathfrak{n}_1'\mathfrak{n}_2'} \mathfrak{n}_5',\, \cdots, \, \sqrt{\mathfrak{n}_1'\mathfrak{n}_2'}  \prod_{i=5}^{n}\mathfrak{n}_i' \right) \nn \\
  (w_{2,1},\, \cdots ,\, w_{2,n-4}) &= \left(\frac{\sqrt{\mathfrak{n}_1'\mathfrak{n}_2'} \mathfrak{n}_5' \cdot w_{5,1}'}{w_{6,1}'},\, \cdots, \, \frac{\sqrt{\mathfrak{n}_1'\mathfrak{n}_2'}  \prod_{i=5}^{n}\mathfrak{n}_i' \cdot w_{n,1}'}{w_{n+1,1}'} \right) \nn \\
  (\mathfrak{n}_1,\, \mathfrak{n}_2,\, q) &=  \left(\frac{q'}{w_{n+1,1}^{\prime 2}},\, w_{n+1,1}',\, \mathfrak{n}_1' \mathfrak{n}_2'\mathfrak{n}_3'\mathfrak{n}_4' \prod_{i=5}^{n+1}\mathfrak{n}_i^{\prime 2}\right)
\end{align}
Replacing the unprimed variables with the primed variables or vice versa, T-duality implies an agreement between the BPS partition functions $\mathcal{I}_{n,1}^A$ and $\mathcal{I}_{n,1}^B$ as explicitly confirmed in the $n=4,5$ cases. It is also straightforward to generalize the fugacity maps for multiple NS5-branes. The detailed analysis would involve understanding the microscopic description of an ON${}^0$ plane.

\section{Conclusion and discussions}
\label{sec:conclusion}

In this work, we studied the $\mathcal{N}=(1,0)$ little string theories that are engineered from type IIA/IIB NS5-branes probing the $D_n$ ALF space.
The string dualities map the NS5-branes and the $D_n$-type ALF space to the NS5-D6-O6 and NS5-ON${}^0$-D6 brane systems, from which we derived 
the 6d effective gauge theories. We have an orthosympletic circular quiver theory on one side, and an affine $\hat{D}_n$ quiver theory on the other side.
Little strings are realized as instanton strings of the effective gauge theories.
Their worldsheet dynamics are described by the two-dimensional $\mathcal{N}=(0,4)$ gauge theories, motivated from various D2-branes introduced to the brane systems.
These 6d/2d gauge theories were used to study the BPS spectra of the little string theories on Omega-deformed $\mathbf{R}^4 \times T^2$.

We utilized the SUSY partition functions of the little string theories to establish their T-duality. 
T-duality is an equivalence between two circle compactified LSTs, interchanging the winding and momentum modes, when their circle radii $R$ and $\tilde{R}$ are related as $\tilde{R} = \alpha'/R$. 
The SUSY partition functions for a T-dual pair of LSTs should agree with each other, after imposing a fugacity map which identifies the winding/momentum fugacities on one side with the momentum/winding fugacities on the other side. 
For those LSTs obtained from an NS5-branes on $D_{n}$ singularities, we found the fugacity map from the associated 5-brane web diagrams and confirmed the agreements between the SUSY partition functions for $n=4, 5$.

By decompactifying either direction in the 5-brane web diagrams, the T-duality is reduced to the duality between the 6d $D_n$ conformal matter theory and the 5d affine $\hat{D}_n$-type quiver gauge theory, or between the 6d ${D}_n$-type quiver gauge theory and the 5d orthosympletic circular quiver theory. This comes from the fact that our LSTs are extensions of those two 6d $\mathcal{N}=(1,0)$ SCFTs. Consequently, our SUSY partition functions also confirmed the 5d/6d dualities studied in \cite{DelZotto:2014hpa,Ohmori:2015pia,Hayashi:2015vhy} as  byproducts. 

One interesting and challenging future direction would be to study the little string theories engineered from type IIA/IIB NS5-branes on the $E_n$-type ALF spaces. 
The effective gauge theories for the type IIB NS5-branes are the affine $\hat{E}_n$ gauge theories according to the Douglas-Moore construction \cite{Douglas1996},
for which most results in Section~\ref{subsec:IIB-index} can be easily extended. However, the effective gauge theories for those type IIA NS5-branes involve several exceptional gauge groups, 
whose instanton strings have not been fully understood yet. \cite{Kim:2016foj,DelZotto:2016pvm,Kim:2017} are some related works on exceptional instanton strings.

\vspace{\baselineskip}
\noindent{\bf\large Acknowledgements}

\noindent
We thank Hee-Cheol Kim, Seok Kim, Sung-Soo Kim, Futoshi Yagi for valuable discussions. We also thank Dario Rosa and especially Sung-Soo Kim for helpful comments on the manuscript.

\vspace{1cm}

\appendix
\section{Explicit expression for $\mathcal{F}_{4,1}$}
\label{sec:1ns5-D4-data}
The Laurant polynomial $\mathcal{F}_{4,1}$ is defined in \eqref{eq:F41} as the numerator of the single-particle index for the LST of a single NS5-brane probing the $D_4$ ALF space. It essentially contains all BPS invariants.
We rearrange it into 
\begin{align}
\mathcal{F}_{4,1} = \sum_{l=0}^\infty \sum_{r=0}^\infty (-1)^{l+r} \chi_{l/2}(u) \chi_{r/2}(t) \cdot \mathfrak{g}_{l,r}
\end{align}
where $\chi_{s}$ denotes an $SU(2)$ character of spin-$s$ representation. Up to the $\mathfrak{n}'_1\mathfrak{n}'_2\mathfrak{n}'_3\mathfrak{n}'_4\mathfrak{n}_5^{\prime 2}$ order,
\begin{align}
\mathfrak{g}_{0,0} &=  w_{5,1}\big(\mathfrak{n}_1 \mathfrak{n}_5^2+8 \mathfrak{n}_1 \mathfrak{n}_2 \mathfrak{n}_5^2+\mathfrak{n}_2 \mathfrak{n}_5^2+8 \mathfrak{n}_1
   \mathfrak{n}_3 \mathfrak{n}_5^2+32 \mathfrak{n}_1 \mathfrak{n}_2 \mathfrak{n}_3 \mathfrak{n}_5^2+8 \mathfrak{n}_2 \mathfrak{n}_3
   \mathfrak{n}_5^2+\mathfrak{n}_3 \mathfrak{n}_5^2+8 \mathfrak{n}_1 \mathfrak{n}_4 \mathfrak{n}_5^2+32 \mathfrak{n}_1 \mathfrak{n}_2
   \mathfrak{n}_4 \mathfrak{n}_5^2 \nn \\&
   +8 \mathfrak{n}_2 \mathfrak{n}_4 \mathfrak{n}_5^2+32 \mathfrak{n}_1 \mathfrak{n}_3 \mathfrak{n}_4
   \mathfrak{n}_5^2+96 \mathfrak{n}_1 \mathfrak{n}_2 \mathfrak{n}_3 \mathfrak{n}_4 \mathfrak{n}_5^2+32 \mathfrak{n}_2 \mathfrak{n}_3
   \mathfrak{n}_4 \mathfrak{n}_5^2+8 \mathfrak{n}_3 \mathfrak{n}_4 \mathfrak{n}_5^2+\mathfrak{n}_4 \mathfrak{n}_5^2+8 \mathfrak{n}_1
   \mathfrak{n}_5+8 \mathfrak{n}_1 \mathfrak{n}_2 \mathfrak{n}_5+8 \mathfrak{n}_2 \mathfrak{n}_5
   \nn \\&+8 \mathfrak{n}_1 \mathfrak{n}_3
   \mathfrak{n}_5+8 \mathfrak{n}_1 \mathfrak{n}_2 \mathfrak{n}_3 \mathfrak{n}_5+8 \mathfrak{n}_2 \mathfrak{n}_3 \mathfrak{n}_5+8
   \mathfrak{n}_3 \mathfrak{n}_5+8 \mathfrak{n}_1 \mathfrak{n}_4 \mathfrak{n}_5+8 \mathfrak{n}_1 \mathfrak{n}_2 \mathfrak{n}_4
   \mathfrak{n}_5+8 \mathfrak{n}_2 \mathfrak{n}_4 \mathfrak{n}_5+8 \mathfrak{n}_1 \mathfrak{n}_3 \mathfrak{n}_4 \mathfrak{n}_5+8
   \mathfrak{n}_1 \mathfrak{n}_2 \mathfrak{n}_3 \mathfrak{n}_4 \mathfrak{n}_5
   \nn \\&+8 \mathfrak{n}_2 \mathfrak{n}_3 \mathfrak{n}_4 \mathfrak{n}_5+8
   \mathfrak{n}_3 \mathfrak{n}_4 \mathfrak{n}_5+8 \mathfrak{n}_4 \mathfrak{n}_5+8
   \mathfrak{n}_5+\mathfrak{n}_1+\mathfrak{n}_2+\mathfrak{n}_3+\mathfrak{n}_4
   \big)  + qw_{5,1}^{-1} \big(10 \mathfrak{n}_1 \mathfrak{n}_2 \mathfrak{n}_5^2+10 \mathfrak{n}_1 \mathfrak{n}_3 \mathfrak{n}_5^2 \nn \\&
   +62 \mathfrak{n}_1 \mathfrak{n}_2
   \mathfrak{n}_3 \mathfrak{n}_5^2+10 \mathfrak{n}_2 \mathfrak{n}_3 \mathfrak{n}_5^2+10 \mathfrak{n}_1 \mathfrak{n}_4 \mathfrak{n}_5^2+62
   \mathfrak{n}_1 \mathfrak{n}_2 \mathfrak{n}_4 \mathfrak{n}_5^2+10 \mathfrak{n}_2 \mathfrak{n}_4 \mathfrak{n}_5^2+62 \mathfrak{n}_1
   \mathfrak{n}_3 \mathfrak{n}_4 \mathfrak{n}_5^2+272 \mathfrak{n}_1 \mathfrak{n}_2 \mathfrak{n}_3 \mathfrak{n}_4 \mathfrak{n}_5^2\nn
   \\&+62
   \mathfrak{n}_2 \mathfrak{n}_3 \mathfrak{n}_4 \mathfrak{n}_5^2+10 \mathfrak{n}_3 \mathfrak{n}_4 \mathfrak{n}_5^2+9 \mathfrak{n}_1
   \mathfrak{n}_5+18 \mathfrak{n}_1 \mathfrak{n}_2 \mathfrak{n}_5+9 \mathfrak{n}_2 \mathfrak{n}_5+18 \mathfrak{n}_1 \mathfrak{n}_3
   \mathfrak{n}_5+27 \mathfrak{n}_1 \mathfrak{n}_2 \mathfrak{n}_3 \mathfrak{n}_5+18 \mathfrak{n}_2 \mathfrak{n}_3 \mathfrak{n}_5\nn
   \\&+9
   \mathfrak{n}_3 \mathfrak{n}_5+18 \mathfrak{n}_1 \mathfrak{n}_4 \mathfrak{n}_5+27 \mathfrak{n}_1 \mathfrak{n}_2 \mathfrak{n}_4
   \mathfrak{n}_5
   +18 \mathfrak{n}_2 \mathfrak{n}_4 \mathfrak{n}_5+27 \mathfrak{n}_1 \mathfrak{n}_3 \mathfrak{n}_4 \mathfrak{n}_5+36
   \mathfrak{n}_1 \mathfrak{n}_2 \mathfrak{n}_3 \mathfrak{n}_4 \mathfrak{n}_5+27 \mathfrak{n}_2 \mathfrak{n}_3 \mathfrak{n}_4
   \mathfrak{n}_5 \nn
   \\&+18 \mathfrak{n}_3 \mathfrak{n}_4 \mathfrak{n}_5+9 \mathfrak{n}_4
   \mathfrak{n}_5+\mathfrak{n}_1+\mathfrak{n}_2+\mathfrak{n}_3+\mathfrak{n}_4+8\big)\\
\mathfrak{g}_{1,0} &= \mathfrak{n}_1 \mathfrak{n}_2 \mathfrak{n}_3 \mathfrak{n}_4 \mathfrak{n}_5^2 w_{5,1}^2+2 \mathfrak{n}_1 \mathfrak{n}_2 \mathfrak{n}_3
   \mathfrak{n}_4 \mathfrak{n}_5^2 + q \big(\mathfrak{n}_1 \mathfrak{n}_5^2+28 \mathfrak{n}_1 \mathfrak{n}_2 \mathfrak{n}_5^2+\mathfrak{n}_2 \mathfrak{n}_5^2+28 \mathfrak{n}_1
   \mathfrak{n}_3 \mathfrak{n}_5^2+181 \mathfrak{n}_1 \mathfrak{n}_2 \mathfrak{n}_3 \mathfrak{n}_5^2+28 \mathfrak{n}_2 \mathfrak{n}_3
   \mathfrak{n}_5^2 \nn \\&
   +\mathfrak{n}_3 \mathfrak{n}_5^2+28 \mathfrak{n}_1 \mathfrak{n}_4 \mathfrak{n}_5^2+181 \mathfrak{n}_1 \mathfrak{n}_2
   \mathfrak{n}_4 \mathfrak{n}_5^2+28 \mathfrak{n}_2 \mathfrak{n}_4 \mathfrak{n}_5^2+181 \mathfrak{n}_1 \mathfrak{n}_3 \mathfrak{n}_4
   \mathfrak{n}_5^2+798 \mathfrak{n}_1 \mathfrak{n}_2 \mathfrak{n}_3 \mathfrak{n}_4 \mathfrak{n}_5^2+181 \mathfrak{n}_2 \mathfrak{n}_3
   \mathfrak{n}_4 \mathfrak{n}_5^2 \nn \\&
   +28 \mathfrak{n}_3 \mathfrak{n}_4 \mathfrak{n}_5^2+\mathfrak{n}_4 \mathfrak{n}_5^2+10 \mathfrak{n}_1
   \mathfrak{n}_5+19 \mathfrak{n}_1 \mathfrak{n}_2 \mathfrak{n}_5+10 \mathfrak{n}_2 \mathfrak{n}_5+19 \mathfrak{n}_1 \mathfrak{n}_3
   \mathfrak{n}_5+28 \mathfrak{n}_1 \mathfrak{n}_2 \mathfrak{n}_3 \mathfrak{n}_5+19 \mathfrak{n}_2 \mathfrak{n}_3 \mathfrak{n}_5+10
   \mathfrak{n}_3 \mathfrak{n}_5 \nn \\&
   +19 \mathfrak{n}_1 \mathfrak{n}_4 \mathfrak{n}_5+28 \mathfrak{n}_1 \mathfrak{n}_2 \mathfrak{n}_4
   \mathfrak{n}_5+19 \mathfrak{n}_2 \mathfrak{n}_4 \mathfrak{n}_5+28 \mathfrak{n}_1 \mathfrak{n}_3 \mathfrak{n}_4 \mathfrak{n}_5+37
   \mathfrak{n}_1 \mathfrak{n}_2 \mathfrak{n}_3 \mathfrak{n}_4 \mathfrak{n}_5+28 \mathfrak{n}_2 \mathfrak{n}_3 \mathfrak{n}_4
   \mathfrak{n}_5+19 \mathfrak{n}_3 \mathfrak{n}_4 \mathfrak{n}_5+10 \mathfrak{n}_4
   \mathfrak{n}_5\nn 
   \\&+\mathfrak{n}_5+\mathfrak{n}_1+\mathfrak{n}_2+\mathfrak{n}_3+\mathfrak{n}_4\big) + q w_{5,1}^{-2} \mathfrak{n}_1 \mathfrak{n}_2 \mathfrak{n}_3 \mathfrak{n}_4 \mathfrak{n}_5^2\\
\mathfrak{g}_{0,1} &= \mathfrak{n}_1 \mathfrak{n}_2 \mathfrak{n}_3 \mathfrak{n}_5^2+\mathfrak{n}_1 \mathfrak{n}_2 \mathfrak{n}_4 \mathfrak{n}_5^2+\mathfrak{n}_1
   \mathfrak{n}_3 \mathfrak{n}_4 \mathfrak{n}_5^2+4 \mathfrak{n}_1 \mathfrak{n}_2 \mathfrak{n}_3 \mathfrak{n}_4
   \mathfrak{n}_5^2+\mathfrak{n}_2 \mathfrak{n}_3 \mathfrak{n}_4 \mathfrak{n}_5^2+\mathfrak{n}_1 \mathfrak{n}_5+\mathfrak{n}_1 \mathfrak{n}_2
   \mathfrak{n}_5+\mathfrak{n}_2 \mathfrak{n}_5+\mathfrak{n}_1 \mathfrak{n}_3 \mathfrak{n}_5 \nn\\&
   +\mathfrak{n}_1 \mathfrak{n}_2 \mathfrak{n}_3
   \mathfrak{n}_5+\mathfrak{n}_2 \mathfrak{n}_3 \mathfrak{n}_5+\mathfrak{n}_3 \mathfrak{n}_5+\mathfrak{n}_1 \mathfrak{n}_4
   \mathfrak{n}_5+\mathfrak{n}_1 \mathfrak{n}_2 \mathfrak{n}_4 \mathfrak{n}_5+\mathfrak{n}_2 \mathfrak{n}_4 \mathfrak{n}_5+\mathfrak{n}_1
   \mathfrak{n}_3 \mathfrak{n}_4 \mathfrak{n}_5+\mathfrak{n}_1 \mathfrak{n}_2 \mathfrak{n}_3 \mathfrak{n}_4 \mathfrak{n}_5+\mathfrak{n}_2
   \mathfrak{n}_3 \mathfrak{n}_4 \mathfrak{n}_5
   \nn\\&+\mathfrak{n}_3 \mathfrak{n}_4 \mathfrak{n}_5+\mathfrak{n}_4
   \mathfrak{n}_5+\mathfrak{n}_5+\mathfrak{n}_1+\mathfrak{n}_2+\mathfrak{n}_3+\mathfrak{n}_4 + w_{5,1}^2 \big(8 \mathfrak{n}_1 \mathfrak{n}_5^2+31 \mathfrak{n}_1 \mathfrak{n}_2 \mathfrak{n}_5^2+8 \mathfrak{n}_2 \mathfrak{n}_5^2+31 \mathfrak{n}_1
   \mathfrak{n}_3 \mathfrak{n}_5^2+88 \mathfrak{n}_1 \mathfrak{n}_2 \mathfrak{n}_3 \mathfrak{n}_5^2\nn\\&
   +31 \mathfrak{n}_2 \mathfrak{n}_3   \mathfrak{n}_5^2+8 \mathfrak{n}_3 \mathfrak{n}_5^2
   +31 \mathfrak{n}_1 \mathfrak{n}_4 \mathfrak{n}_5^2+88 \mathfrak{n}_1 \mathfrak{n}_2
   \mathfrak{n}_4 \mathfrak{n}_5^2   +31 \mathfrak{n}_2 \mathfrak{n}_4 \mathfrak{n}_5^2+88 \mathfrak{n}_1 \mathfrak{n}_3 \mathfrak{n}_4
   \mathfrak{n}_5^2+222 \mathfrak{n}_1 \mathfrak{n}_2 \mathfrak{n}_3 \mathfrak{n}_4 \mathfrak{n}_5^2+88 \mathfrak{n}_2 \mathfrak{n}_3
   \mathfrak{n}_4 \mathfrak{n}_5^2\nn\\&
   +31 \mathfrak{n}_3 \mathfrak{n}_4 \mathfrak{n}_5^2+8 \mathfrak{n}_4 \mathfrak{n}_5^2+\mathfrak{n}_5^2+8
   \mathfrak{n}_1 \mathfrak{n}_5+8 \mathfrak{n}_1 \mathfrak{n}_2 \mathfrak{n}_5+8 \mathfrak{n}_2 \mathfrak{n}_5+8 \mathfrak{n}_1
   \mathfrak{n}_3 \mathfrak{n}_5+8 \mathfrak{n}_1 \mathfrak{n}_2 \mathfrak{n}_3 \mathfrak{n}_5+8 \mathfrak{n}_2 \mathfrak{n}_3
   \mathfrak{n}_5+8 \mathfrak{n}_3 \mathfrak{n}_5
   \nn\\&
   +8 \mathfrak{n}_1 \mathfrak{n}_4 \mathfrak{n}_5+8 \mathfrak{n}_1 \mathfrak{n}_2
   \mathfrak{n}_4 \mathfrak{n}_5+8 \mathfrak{n}_2 \mathfrak{n}_4 \mathfrak{n}_5+8 \mathfrak{n}_1 \mathfrak{n}_3 \mathfrak{n}_4
   \mathfrak{n}_5+8 \mathfrak{n}_1 \mathfrak{n}_2 \mathfrak{n}_3 \mathfrak{n}_4 \mathfrak{n}_5+8 \mathfrak{n}_2 \mathfrak{n}_3 \mathfrak{n}_4
   \mathfrak{n}_5+8 \mathfrak{n}_3 \mathfrak{n}_4 \mathfrak{n}_5+8 \mathfrak{n}_4 \mathfrak{n}_5+8 \mathfrak{n}_5+1\big)\nn \\&
   + qw_{5,1}^{-2} \big( \mathfrak{n}_1 \mathfrak{n}_2 \mathfrak{n}_5^2+\mathfrak{n}_1 \mathfrak{n}_3 \mathfrak{n}_5^2+4 \mathfrak{n}_1 \mathfrak{n}_2 \mathfrak{n}_3
   \mathfrak{n}_5^2+\mathfrak{n}_2 \mathfrak{n}_3 \mathfrak{n}_5^2+\mathfrak{n}_1 \mathfrak{n}_4 \mathfrak{n}_5^2+4 \mathfrak{n}_1
   \mathfrak{n}_2 \mathfrak{n}_4 \mathfrak{n}_5^2+\mathfrak{n}_2 \mathfrak{n}_4 \mathfrak{n}_5^2+4 \mathfrak{n}_1 \mathfrak{n}_3
   \mathfrak{n}_4 \mathfrak{n}_5^2
      \nn\\&
      +15 \mathfrak{n}_1 \mathfrak{n}_2 \mathfrak{n}_3 \mathfrak{n}_4 \mathfrak{n}_5^2+4 \mathfrak{n}_2
   \mathfrak{n}_3 \mathfrak{n}_4 \mathfrak{n}_5^2+\mathfrak{n}_3 \mathfrak{n}_4 \mathfrak{n}_5^2+\mathfrak{n}_1 \mathfrak{n}_5+2
   \mathfrak{n}_1 \mathfrak{n}_2 \mathfrak{n}_5+\mathfrak{n}_2 \mathfrak{n}_5+2 \mathfrak{n}_1 \mathfrak{n}_3 \mathfrak{n}_5+3 \mathfrak{n}_1
   \mathfrak{n}_2 \mathfrak{n}_3 \mathfrak{n}_5+2 \mathfrak{n}_2 \mathfrak{n}_3 \mathfrak{n}_5
      \nn\\&
      +\mathfrak{n}_3 \mathfrak{n}_5+2 \mathfrak{n}_1
   \mathfrak{n}_4 \mathfrak{n}_5+3 \mathfrak{n}_1 \mathfrak{n}_2 \mathfrak{n}_4 \mathfrak{n}_5+2 \mathfrak{n}_2 \mathfrak{n}_4
   \mathfrak{n}_5+3 \mathfrak{n}_1 \mathfrak{n}_3 \mathfrak{n}_4 \mathfrak{n}_5+4 \mathfrak{n}_1 \mathfrak{n}_2 \mathfrak{n}_3 \mathfrak{n}_4
   \mathfrak{n}_5+3 \mathfrak{n}_2 \mathfrak{n}_3 \mathfrak{n}_4 \mathfrak{n}_5+2 \mathfrak{n}_3 \mathfrak{n}_4 \mathfrak{n}_5+\mathfrak{n}_4
   \mathfrak{n}_5
      \nn\\&
      +\mathfrak{n}_1+\mathfrak{n}_2+\mathfrak{n}_3+\mathfrak{n}_4+1 \big) + q \big( 47 \mathfrak{n}_1 \mathfrak{n}_5^2+208 \mathfrak{n}_1 \mathfrak{n}_2 \mathfrak{n}_5^2+47 \mathfrak{n}_2 \mathfrak{n}_5^2+208 \mathfrak{n}_1
   \mathfrak{n}_3 \mathfrak{n}_5^2+696 \mathfrak{n}_1 \mathfrak{n}_2 \mathfrak{n}_3 \mathfrak{n}_5^2
   +208 \mathfrak{n}_2 \mathfrak{n}_3
   \mathfrak{n}_5^2
   \nn\\&
   +47 \mathfrak{n}_3 \mathfrak{n}_5^2+208 \mathfrak{n}_1 \mathfrak{n}_4 \mathfrak{n}_5^2+696 \mathfrak{n}_1 \mathfrak{n}_2
   \mathfrak{n}_4 \mathfrak{n}_5^2+208 \mathfrak{n}_2 \mathfrak{n}_4 \mathfrak{n}_5^2+696 \mathfrak{n}_1 \mathfrak{n}_3 \mathfrak{n}_4
   \mathfrak{n}_5^2+2065 \mathfrak{n}_1 \mathfrak{n}_2 \mathfrak{n}_3 \mathfrak{n}_4 \mathfrak{n}_5^2+696 \mathfrak{n}_2 \mathfrak{n}_3
   \mathfrak{n}_4 \mathfrak{n}_5^2
   \nn\\&
   +208 \mathfrak{n}_3 \mathfrak{n}_4 \mathfrak{n}_5^2+47 \mathfrak{n}_4 \mathfrak{n}_5^2+\mathfrak{n}_5^2+48
   \mathfrak{n}_1 \mathfrak{n}_5+66 \mathfrak{n}_1 \mathfrak{n}_2 \mathfrak{n}_5+48 \mathfrak{n}_2 \mathfrak{n}_5+66 \mathfrak{n}_1
   \mathfrak{n}_3 \mathfrak{n}_5+84 \mathfrak{n}_1 \mathfrak{n}_2 \mathfrak{n}_3 \mathfrak{n}_5+66 \mathfrak{n}_2 \mathfrak{n}_3
   \mathfrak{n}_5\nn\\&
   +48 \mathfrak{n}_3 \mathfrak{n}_5+66 \mathfrak{n}_1 \mathfrak{n}_4 \mathfrak{n}_5+84 \mathfrak{n}_1 \mathfrak{n}_2
   \mathfrak{n}_4 \mathfrak{n}_5
   +66 \mathfrak{n}_2 \mathfrak{n}_4 \mathfrak{n}_5+84 \mathfrak{n}_1 \mathfrak{n}_3 \mathfrak{n}_4
   \mathfrak{n}_5+102 \mathfrak{n}_1 \mathfrak{n}_2 \mathfrak{n}_3 \mathfrak{n}_4 \mathfrak{n}_5+84 \mathfrak{n}_2 \mathfrak{n}_3
   \mathfrak{n}_4 \mathfrak{n}_5
   \nn\\&+66 \mathfrak{n}_3 \mathfrak{n}_4 \mathfrak{n}_5+48 \mathfrak{n}_4 \mathfrak{n}_5+30 \mathfrak{n}_5+2
   \mathfrak{n}_1+2 \mathfrak{n}_2+2 \mathfrak{n}_3+2 \mathfrak{n}_4 \big)
\end{align}
where all primes are dropped for simplicity. See the attached Mathematica files for more informations.


\providecommand{\href}[2]{#2}\begingroup\raggedright\endgroup

\end{document}